\documentclass[12pt, a4paper]{article}

\usepackage{amssymb}
\usepackage{amsmath}
\usepackage{amsfonts}
\usepackage{graphicx}
\usepackage{mathrsfs}
\usepackage{comment}
\usepackage{cite}
\def\->{\rightarrow}
\def\<-{\leftarrow}

\newcommand{\Slash}[1]{{\ooalign{\hfil#1\hfil\crcr\raise.167ex\hbox{/}}}}

\newcommand{\beq}{\begin{equation}}  \newcommand{\eeq}{\end{equation}}
\newcommand{\bef}{\begin{figure}}  \newcommand{\eef}{\end{figure}}
\newcommand{\bec}{\begin{center}}  \newcommand{\eec}{\end{center}}
  
\newcommand{\laq}[1]{\label{eq:#1}}  

\newcommand{\Eq}[1]{Eq.(\ref{eq:#1})}

\newcommand{\eq}[1]{(\ref{eq:#1})}

\newcommand{\ab}[1]{\left|{#1}\right|}

\newcommand{\lac}[1]{\label{chap:#1}}

\newcommand{\bed}{\begin{description} \item}
\newcommand{\eed}{\end{description}}
\newcommand{\Fig}[1]{Fig.\ref{fig:#1}}  
\newcommand{\laf}[1]{\label{fig:#1}}
\def\({\left(}
\def\){\right)}

\def\a{\alpha}

\def\f{\phi}
\def\g{\gamma}
\def\h{\theta}

\def\s{\sigma}

\def\tl{\tilde}
\def\*{\dagger}

\newcommand{\AND}{~{\rm and}~}
\newcommand{\EV}{ {\rm eV} }

\newcommand{\GEV}{ {\rm GeV} }

\renewcommand{\thefootnote}{\fnsymbol{footnote}}

\begin{document}
\begin{titlepage}
\begin{center}

\hfill TU-1039,  IPMU17-0031\\

\vskip .75in

{\Large\bf The ALP miracle: unified inflaton and dark matter
}

\vskip .75in

{ \large Ryuji Daido\,$^{a}$\footnote{email: daido@tuhep.phys.tohoku.ac.jp},
    Fuminobu  Takahashi\,$^{a,b}$\footnote{email: fumi@tuhep.phys.tohoku.ac.jp},
    Wen Yin\,$^{c}$\footnote{email: wyin@ihep.ac.cn}}

\vskip 0.25in

\begin{tabular}{ll}
$^{a}$ &\!\! {\em Department of Physics, Tohoku University, }\\
& {\em Sendai, Miyagi 980-8578, Japan}\\[.3em]
$^{b}$ &\!\! {\em Kavli IPMU (WPI), UTIAS,}\\
&{\em The University of Tokyo,  Kashiwa, Chiba 277-8583, Japan}\\[.3em]
$^{c}$ &\!\! {\em IHEP, Chinese Academy of Sciences, Beijing 100049,  China}\\
& {\em }

\end{tabular}

\end{center}
\vskip .5in

\begin{abstract}
We propose a scenario where both inflation and dark matter are described
by a single axion-like particle (ALP) in a unified manner. In a class of the  minimal axion hilltop inflation,
the effective masses at the maximum and mimimum of the potential  have equal 
magnitude but opposite sign, so that the ALP inflaton is light both during inflation and in the true vacuum.
After inflation, most of the ALPs decay and evaporate
into plasma through a coupling to photons, and the remaining ones become dark matter.
We find that the observed CMB and matter power spectrum
as well as the dark matter abundance point to an ALP of mass 
$m_\phi = {\cal O}(0.01)$\,eV and the axion-photon coupling $g_{\phi \gamma \gamma} =
 {\cal O}(10^{-11})$\,GeV$^{-1}$: {\it the ALP miracle}. The suggested parameter region is within the reach 
 of the next generation axion helioscope, IAXO, and high-intensity laser experiments in the future.
Furthermore,  thermalized ALPs contribute to hot dark matter and its abundance is
given in terms of the effective number of extra neutrino species, $\Delta N_{\rm eff} \simeq 0.03$, 
which can be tested by the future CMB and BAO observations.
We also discuss a case with multiple ALPs, where the coupling
to photons can be enhanced in the early Universe by an order of magnitude or more, which enlarges
the parameter space for the ALP miracle. The heavy ALP { plays a role of the waterfall field in hybrid inflation,
 and reheats the Universe, }and it can be  searched for in
various experiments such as SHiP.
\end{abstract}

\end{titlepage}
\setcounter{footnote}{0}
\setcounter{page}{1}

\renewcommand{\thefootnote}{\arabic{footnote}}
\section{Introduction}
The slow-roll inflation paradigm has been established by the precise measurements
of temperature and polarization anisotropies of the cosmic microwave background 
radiation (CMB)~\cite{Ade:2015lrj}. For successful inflation, the inflaton potential needs to be sufficiently flat,
and such a flat potential can be naturally realized if the inflaton is an axion. This is because 
the axion enjoys a shift symmetry, which keeps the potential flat at the perturbative level.
While the axion potential is generated by various non-perturbative effects, it is still under
control if the discrete shift symmetry remains unbroken, i.e., if the potential is periodic. 
A periodic potential can be expanded in Fourier series and given by a sum of cosine functions.
Of course, a generic periodic potential does not necessarily lead to successful slow-roll inflation;
it must be sufficiently flat over a certain field range. 

The natural inflation is the simplest axion inflation where the inflaton potential
consists of a single cosine function~\cite{Freese:1990rb,Adams:1992bn},
\begin{align}
 \Lambda^4 \left(1 - \cos\(\phi \over f \)\right),
\end{align}
where  $f$ is the decay constant, and $\Lambda$ determines the inflation scale.
For successful inflation, however, $f$ is required to be about five times larger than the reduced Planck mass,
$f \gtrsim 5 M_{pl}$~\cite{Ade:2015lrj}. Such a large decay constant has been questioned from the perspective
 of a UV theory including gravity. See e.g. Refs.~\cite{Rudelius:2014wla,delaFuente:2014aca,
Rudelius:2015xta,Montero:2015ofa,Brown:2015iha}  for discussion based on the 
weak gravity conjecture~\cite{ArkaniHamed:2006dz}. 

A simple extension of the natural inflation is a multi-natural 
inflation~\cite{Czerny:2014wza,Czerny:2014xja,Czerny:2014qqa}, 
where the axion potential receives multiple sinusoidal functions with different height and period.\footnote{
The axion in a potential with multiple sinusoidal terms was also discussed 
in a context of dark matter~\cite{Higaki:2016yqk,Jaeckel:2016qjp} and curvaton~\cite{Takahashi:2013tj}.
}
In particular,  a small-field axion hilltop inflation model was first realized in this context. 
Since then the axion hilltop inflation and its variants 
have been studied in various string-inspired set-up~\cite{Czerny:2014xja,Czerny:2014qqa,Higaki:2014sja,
Croon:2014dma,Higaki:2015kta,Choi:2015aem,Higaki:2016ydn}.
 In particular, such axion hilltop inflation nicely fits into a scheme of the axion 
 landscape~\cite{Higaki:2014pja,Higaki:2014mwa} where many axions with mass and kinetic
 mixings receive various shift symmetry breaking potentials with a complicated landscape.

In the minimal axion hilltop inflation, the inflaton potential consists of
two sinusoidal functions~\cite{Czerny:2014wza}:
\begin{align}
\label{DIV} 
V_{\rm inf}(\phi) = \Lambda^4\(\cos\({\f \over f } + \theta \)-{\kappa \over n^2}\cos\({n\f \over f }\)\)+{\rm const.},
\end{align}
where $n (>1)$ is a rational number, $\kappa$ is a numerical coefficient, $\theta$ is a relative phase, and 
the last term is a constant required to realize the vanishingly small cosmological constant in the present vacuum.
For the moment we take $\theta = 0$ and $\kappa = 1$. Then,
successful inflation takes place in the vicinity of the origin  where the curvature, or the effective mass, is much smaller than the Hubble parameter
during inflation. Interestingly, if $n$ is an odd integer,  the axion mass at the potential maximum  and minimum is
equal in magnitude but has an opposite sign. In other words, the axion has a flat-top and flat-bottomed potential
(see Fig.~\ref{fig:pot}). This interesting feature of the axion potential is partly due to the unbroken 
discrete shift symmetry, which requires a periodic axion potential.\footnote{
This should be contrasted to many other inflation models where the inflaton
potential during inflation is not directly related to that around the potential minimum; in this case,
it is impossible to probe directly the inflaton properties during inflation by any far-future experiments, which are only able to 
study the inflaton at the potential minimum.} As a result, the inflaton (axion) is massless at the potential minimum.
In fact, as we shall see later, if a non-zero phase $\theta$ or deviation of $\kappa$ from unity  is introduced to explain 
the observed spectral index, a tiny axion mass, $m_\phi$, is generated at the potential minimum. 
The relation between the axion mass $m_\phi$ and decay constant $f$ 
is then fixed by the Planck normalization of  curvature perturbations and the spectral index.
Since the inflaton remains light in the present vacuum, it may be long-lived on a cosmological time scale and
contribute to dark matter.\footnote{The possibility that the remnant inflaton condensate due to incomplete reheating 
becomes dark matter dates back to the seminal papers on preheating~\cite{Kofman:1994rk,Kofman:1997yn}. 
See Refs.~\cite{Mukaida:2014kpa, Bastero-Gil:2015lga} for recent works along the same line taking account of 
dissipation effects. Also, thermalized inflaton particles may become WIMP dark 
matter~\cite{Lerner:2009xg,Okada:2010jd,Khoze:2013uia,Bastero-Gil:2015lga}.
A possibility of inflatino dark matter was pointed out in Ref.~\cite{Nakayama:2010kt}.
}

Let us assume that the axion is coupled to photons for successful reheating, and
such an axion  is called the axion-like particle (ALP). Then, the ALP mass, $m_\phi$,
and its coupling to photons, $g_{\phi \gamma \gamma}$,  are fixed by the requirement of successful inflation, 
up to a model-dependent numerical factor of ${\cal O}(1)$. After inflation, while most of the ALPs decay and evaporate 
into  radiation through the coupling to photons, some of them remain and serve as (decaying) dark matter if long-lived. 
As we shall see, the ALP mass $m_\phi$ is constrained to be heavier than ${\cal O}(10^{-2})$\,eV
by the small-scale matter power spectrum, which, in turn, implies that only ${\cal O}(1)$\% or less of the initial ALPs
could contribute to dark matter and the rest of it should evaporate into plasma. This sets a lower bound on
the dissipation rate, or equivalently, on $g_{\phi \gamma \gamma}$. Taking account of the uncertainty in the
order-of-magnitude estimate of the dissipation rate as well as possible couplings to weak gauge bosons, 
the above scenario 
points to  $m_\phi = {\cal O}(0.01)$\,eV and  $g_{\phi \gamma \gamma} = {\cal O}(10^{-11})$\,GeV$^{-1}$,
which are within the reach of the next generation axion helioscope IAXO~\cite{Irastorza:2011gs,Armengaud:2014gea} and  the proposed purely laser-based stimulated photon-photon collider \cite{Fujii:2010is,Homma:2017cpa}.

This is a non-trivial coincidence because three independent conditions, i.e., successful inflation (normalization of curvature perturbation 
and the spectral index), successful reheating, and the observed dark matter abundance 
meet at one point. We call such a coincidence as {\it the ALP miracle}.\footnote{
In this sense, the ALP miracle is not parallel to the WIMP miracle which is about the coincidence
in explaining the dark matter abundance with  new physics around the weak scale. 
} 

We will also discuss a case with multiple ALPs and show that the ALP miracle still holds.
Once a certain combination of the ALPs is identified with the inflaton, the slow-roll inflation dynamics is same as
the case with a single ALP, but the dynamics after inflation is slightly more involved.  
Due to the mixing between ALPs, the coupling of the inflaton to photons can be enhanced in the early Universe, 
which enlarges the parameter space for the ALP miracle. The heavy ALPs can be
produced after inflation like the waterfall field in hybrid inflation~\cite{Copeland:1994vg,
Dvali:1994ms,Linde:1997sj} or through resonant conversion process through a mixing. 
Thus produced heavy ALPs decay and/or evaporate into the standard model (SM) particles, and reheat the Universe
in some case. Such heavy ALPs and their decay (which may be responsible for the reheating)  can be searched for in
various axion search experiments such as SHiP~\cite{Alekhin:2015byh,Dobrich:2015jyk}. 

The rest of this paper is organized as follows. In Sec.~\ref{sec2} we explain how the ALP miracle is realized
in a model with a single ALP.  In Sec.~\ref{sec3} we explore a scenario with multiple ALPs, 
and study the difference from the single ALP case and  investigate the production mechanism of  the 
heavy ALPs. The last section is devoted to discussion and conclusions.

\section{The ALP miracle}
\label{sec2}

\subsection{Axion Hilltop Inflation}
Let us consider the minimal axion hilltop inflation with the potential (\ref{DIV}), which is one realization of
the multi-natural inflation~\cite{Czerny:2014wza,Czerny:2014xja,Czerny:2014qqa}. 
For the moment let us set $\theta = 0$ and $\kappa = 1$ so that the inflaton mass vanishes at the origin, $\phi = 0$.
In general, the inflaton potential must be very flat for successful slow-roll inflation, 
which determines the prefactor of the second term.  Such flat-top potential is realized in models with 
extra dimensions~\cite{Croon:2014dma} and it is also obtained in a certain limit 
of the elliptic function~\cite{Higaki:2015kta}.

\begin{figure}[!t]
\begin{center}  
   \includegraphics[width=100mm]{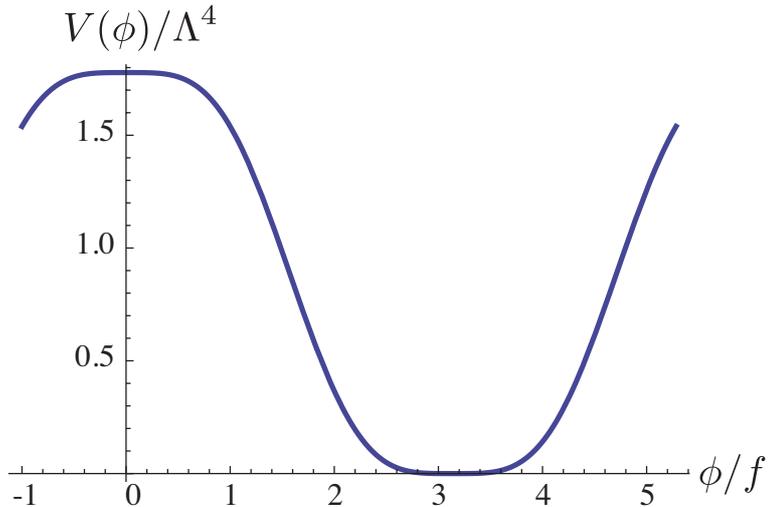}
\end{center}
\caption{
 The axion potential (\ref{DIV}) with $n=3$, and $\h=0$. 
}
\label{fig:pot}
\end{figure}

If $n$ is an odd integer, the potential satisfies
\beq
\laq{inv}
V(\phi + \pi f) = - V(\phi)+ {\rm const.},
\eeq
where the constant term does not depend on $\phi$. 
Interestingly, this implies a massless inflaton at both the maximum $(\phi = 0)$ and minimum $(\phi = \pi f)$ 
of the potential. The potential with $n=3$  is illustrated in Fig.\ref{fig:pot}. 
Throughout this paper we focus on a case with an odd $n$, but discussion on the inflaton
dynamics during inflation is valid for any rational number $n > 1$.  
As we shall see shortly, the inflaton still remains light when we introduce a small nonzero CP phase 
$\theta$ to give a better fit to the observed spectral index. Such light inflaton may be relevant for low-energy physics, and it opens up a possibility that
the inflaton dynamics is directly probed by experiments.

The potential near the origin is approximated by
\beq
\label{Vinf}
V \simeq   V_0-  \lambda {\f^4 } 
+ \cdots,
\eeq
where we have defined
 \begin{align}
 \label{V0}
 V_0 &\equiv  V(0) - V(\pi f) 
 = 2{n^2-1\over n^2} \Lambda^4 \\
 \lambda & \equiv {n^2-1 \over 4!}\({\Lambda\over f}\)^4.
  \label{lambda}
\end{align}
In the second equality of Eq.~(\ref{V0}) we have required that the cosmological constant 
 (almost) vanishes in the present vacuum, $\phi = \pi f$. Thus, the minimal axion hilltop inflation 
 with $\theta = 0$  is equivalent to the hilltop quartic inflation, as far as the dynamics during 
 inflation is concerned.

Now let us estimate the curvature perturbation generated by the inflation. 
According to the Planck and other ground-based experiments,
the spectral index $n_s$  and the tensor-to-scalar ratio $r$ are~\cite{Ade:2015lrj}
\begin{align}
\label{ns}
n_s& =0.968\pm 0.006,\\
r &< 0.11 ~~(95\%{\rm CL}),
\end{align}
and the Planck normalization on the curvature perturbation is given by
\beq
\label{pn}
P_{\mathcal{R}} \simeq 2.2 \times 10^{-9},
\eeq
at the pivot scale $k_0 = 0.002\, {\rm Mpc}^{-1}$.
On the other hand, it is known that the spectral index in the hilltop quartic inflation model tends to be
smaller than the observed value.

The spectral index $n_s$ is given in terms of the slow-roll parameters as
\beq
\laq{nsform}
n_s \simeq 1-6\, \varepsilon(\f_*)+2 \eta(\f_*),
\eeq
where the slow-roll parameters are defined by
\begin{equation}
\laq{defeta}
  \varepsilon(\f) \equiv \frac{M_{pl}^2}{2}\left(\frac{V'}{V}\right)^2 , \quad \eta(\f) \equiv M_{pl}^2\left(\frac{V''}{V}\right).
\end{equation}
Here $M_{pl} \simeq 2.4 \times 10^{18}$\,GeV is the reduced Planck mass, and
the subscript $*$  implies that the variable is evaluated
at the horizon exit of the cosmological scales. The evolution of the inflaton can be expressed 
in terms of the e-folding number $N_*$ by solving
\beq
\laq{xecon}
N_* = \int^{\f_{\rm end}}_{\f_*}{{H \over \dot{\f}} d \f} \simeq \int^{\f_{\rm end}}_{\f_*}{{3H^2 \over -V'} d \f},
\eeq
where $H$ is the Hubble parameter, $\dot{\f}$ is the time derivative of the inflaton field $\f$, 
and we have used the slow-roll equation of motion, $3 H \dot{\phi} + V' \simeq 0$, in the second equality.
We assume that $\dot{\phi} > 0$ during inflation without loss of generality. 
Here $\f_{\rm end}$ is the field value  at the end of inflation defined by $|\eta(\phi_{\rm end})|= 1$, 
and it is given by
\begin{align}
\label{phiend}
\phi_{\rm end} & \simeq \frac{2}{n} \frac{f^2}{M_{pl}} \ll \phi_{\rm min},
\end{align}
where $\phi_{\rm min} = \pi f$ is the field value at the potential minimum. 
Note that
$\varepsilon \ll |\eta|$ generally holds during inflation in the small-field inflation
and  it is $\eta$ that determines the deviation of $n_s$ from unity. 
Substituting the inflaton potential (\ref{Vinf}) into the above equations, one obtains
\beq
\label{nshill}
n_s\simeq 1-{3 \over N_*},
\eeq
whereas the tensor-to-scalar ratio $r$ is negligibly small for the sub-Planckian decay constant, $f \ll M_{pl}$.

The curvature perturbation is given by
\beq
\laq{pln}
P_{\mathcal{R}} = \(\frac{H_*^2}{2\pi \dot{\phi}_*}\)^2 \simeq 
\frac{V(\phi_*)^3}{12 \pi^2 V'(\phi_*)^2 M_{pl}^6}.
\eeq 
Then, the Planck normalization (\ref{pn}) fixes the quartic coupling as
\beq
\label{pninni}
\lambda \simeq  7.5 \times 10^{-14} \({N_* \over 50}\)^{-3}.
\eeq
To evaluate $n_s$ and $\lambda$, therefore, one needs to know the precise value of the e-folding number.

\begin{figure}[!t]
\begin{center}  
   \includegraphics[width=105mm]{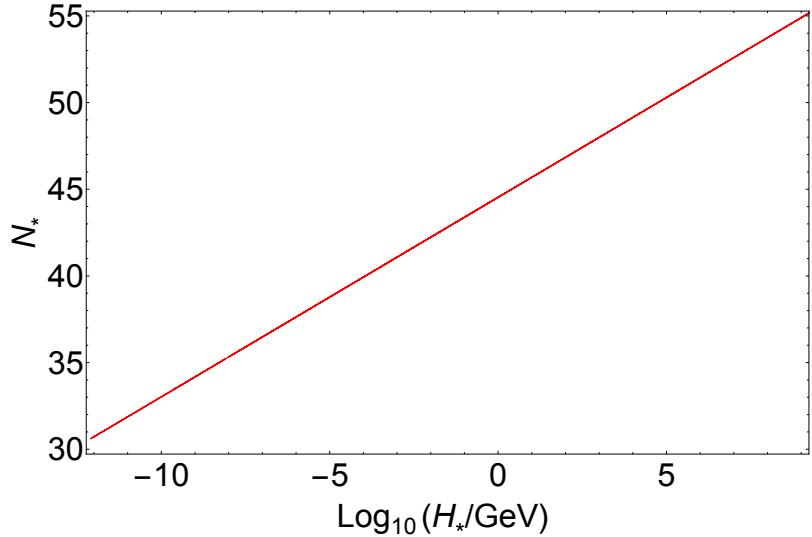}
\end{center}
\caption{
The e-folding number $N_*$ as a function of the Hubble parameter at the horizon exit, $H_*$. 
}
\label{fig:e-folding}
\end{figure}

The e-folding number is fixed once the inflation scale and thermal history after inflation are given. In the model
at hand, the inflaton potential is  quartic around the potential minimum, and so, the inflaton energy density
decreases like radiation after inflation. In this case, the e-folding number is given by
\beq
\laq{efold}
N_*\simeq 61+ \ln \({H_* \over H_{\rm end}}\)^\frac{1}{2}+ \ln{\({H_{\rm end} \over 10^{14} \GEV}\)^\frac{1}{2}},
\eeq
where $H_*$ ($H_{\rm end}$) is the Hubble parameter at the horizon exit (the end of the inflation). 
In the small-field inflation, the Hubble parameter remains almost constant during inflation, and so,
we can simply set $ H_* \simeq H_{\rm end} = H_{\rm inf} $. The relation between $H_*$ and $N_*$ is
shown in Fig.~\ref{fig:e-folding}.
Since $H_{\rm inf}$ is smaller than $10^{14}$\,GeV for a sub-Planckian decay constant, 
the e-folding number cannot exceed $60$. Therefore, the predicted spectral index (\ref{nshill})
must be smaller than $0.95$, which is in a clear tension with the observed value (\ref{ns}).

As shown in Refs.~\cite{Czerny:2014wza, Takahashi:2013cxa}, the spectral index can be increased by introducing 
a small but nonzero CP phase, $\theta$. For sufficiently small $\theta (>0)$,  one can expand the potential around the origin,
\beq
V \simeq V_0 - \lambda \phi^4 - \Lambda^4 \theta \frac{\phi}{f} + \cdots.
\eeq
Thus, the CP phase induces a linear term. While the linear term does not directly contribute to $\eta$, 
it does change the inflaton field value at the horizon exit, $\phi_*$.
This is because, as $|V'|$ increases due to $\theta$, 
$\phi_*$ decreases for the same e-folding number (cf. \Eq{xecon}).
Since $\eta$ depends on $\phi_*$ as
\beq
\laq{eta}
\eta(\f_*)\propto \f_*^2,
\eeq 
one can increase $n_s$ by introducing a small linear term which decreases $\phi_*$.

Let us estimate the inflaton mass induced by the relative phase $\theta$. In the limit of $\theta = 0$,
the inflaton was massless, but a small mass is generated by a nonzero $\theta$. The inflaton mass at the
potential minimum $\phi = \phi_{\rm min}$ is
\beq
m_\phi^2 \equiv V''(\phi_{\rm min}) \simeq \( \left(\frac{9(n^2-1)}{2} \right)^\frac{1}{6} \theta^\frac{1}{3} \frac{\Lambda^2}{f}\)^2,
\eeq
where we have approximated $|\theta| \ll 1$. 
The inflaton mass at the potential maximum is equal to $m_\phi^2$ but with an opposite sign.  
The inflation takes place in the vicinity of the potential maximum, and 
the curvature of the potential at $\phi = \phi_*$ is comparable to $-m_\phi^2$.  In order to increase 
$n_s$ by ${\cal O}(0.01)$, therefore, we need 
\beq
\label{m2H2}
 \frac{m_\phi^2}{H_*^2}  \sim  \frac{|V''(\phi_*)|}{H_*^2}  = {\cal O}(0.01).
\eeq
Thus, even when we introduce a nonzero phase $\theta (\sim f^3/M_{pl}^3)$ to explain the
observed spectral index, the inflaton mass at the potential minimum is
still lower than the Hubble parameter. In particular, it is much smaller than
the typical mass scale  of each cosine function, $\Lambda^2/f$.
We emphasize that the second equality of (\ref{m2H2}) holds generally  
in a small-field inflation. In particular, it is also possible to increase the spectral index by
taking $\kappa \ne 1$, but even in this case, as long as the spectral index is within the observed
value (\ref{ns}), the relation (\ref{m2H2}) should hold. To simplify our analysis we vary only $\theta$
in the following analysis, 
but our argument is not modified even if one also varies $\kappa$. See footnote \ref{f7}
for a more quantitative estimate.

Similarly,  the quartic coupling of the inflaton at the minimum is equal to $\lambda$ fixed by
the Planck normalization:
\beq
\laq{rela2}
V^{(4)}(\f_{\rm min}) \simeq 4! \lambda.
\eeq
It is interesting to note that the quartic coupling at the potential minimum is fixed by
the  Planck normalization of the curvature perturbation and the relation between the inflaton mass
and the decay constant (or $H_*$) is fixed by the spectral index.

\begin{figure}[t!]
\begin{center}  
   \includegraphics[width=105mm]{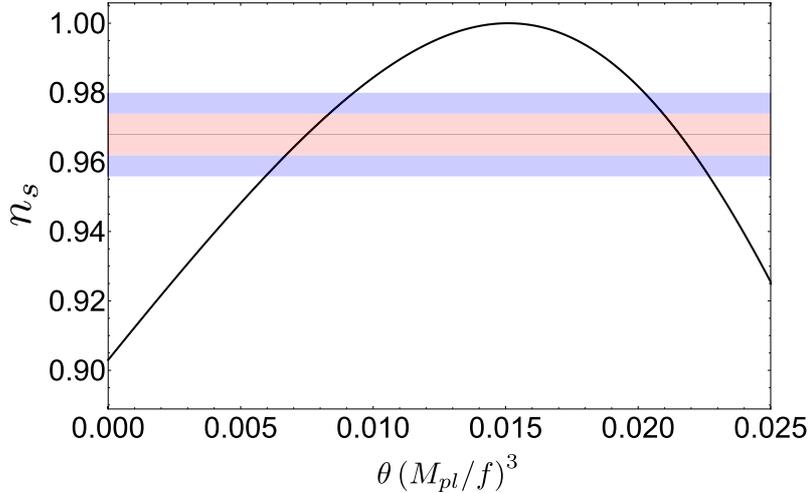}
\end{center}
\caption{The spectral index $n_s$ as a function of the CP phase $\theta$ for ${f/M_{pl}}=10^{-11}$.
The Planck observation (\ref{ns}) is shown as red ($1\sigma$) and blue ($2\sigma$) bands.
}\label{fig:ns}
\end{figure}

We have numerically solved the inflaton dynamics with the potential (\ref{DIV})
to evaluate the spectral index, the quartic coupling, the inflaton mass and the inflation scale.
In Fig.~\ref{fig:ns} we show the spectral index as a function of $\theta$.  As one can see, the spectral
index can be enhanced to be within the observed range (\ref{ns}) for non-zero values of $\theta$, and
in particular, there are two solutions.
This is because the spectral index becomes equal to unity for some value of $\theta$ where $\phi_*\simeq 0$,
and $\eta(\f_*)$ is almost an even function (cf. \Eq{eta}). Thus, there are correspondingly 
two solutions, $\phi_* >0 $ and $\phi_*<0$. 
A larger (smaller) $\theta$ corresponds 
to $\phi_*  < 0$ ($\phi_*  > 0)$, which will be referred to as the solution 1(2) in the following.

\begin{figure}[t!]
\begin{center}  
   \includegraphics[width=105mm]{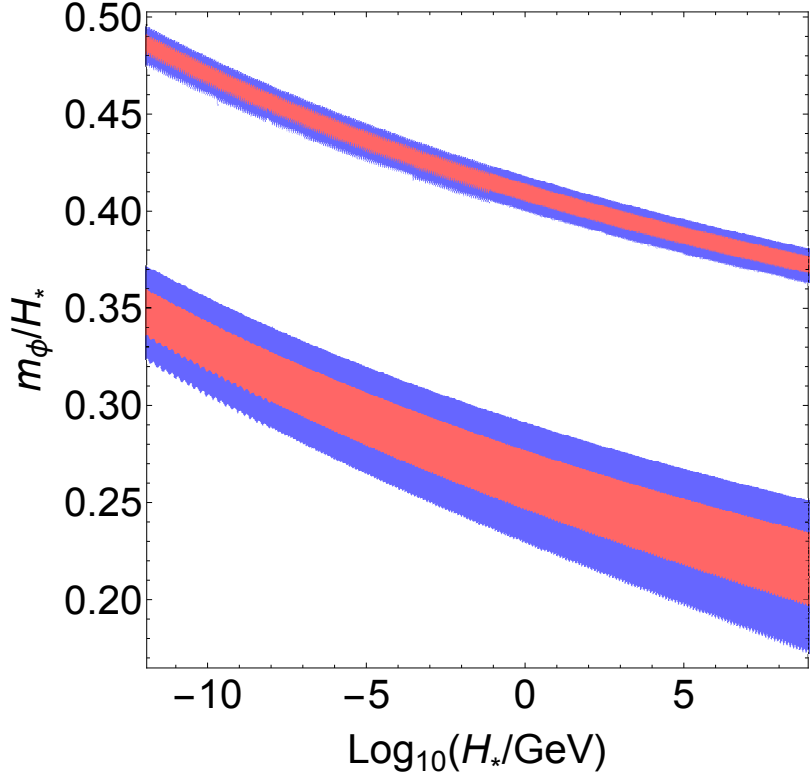}
\end{center}
\caption{ The ratio of the inflaton mass, $m_\f$,  to the Hubble parameter at the horizon exit, $H_*$,
as a function of $H_*$. 
The red (blue) band represents the parameter region where the spectral index explains the observed
value within $1\s~ (2\s)$ CL.}
\label{fig:m}
\end{figure}

 In Fig.~\ref{fig:m} we show
the ratio of the inflaton mass to the Hubble parameter, $m_\f  /H_*$,  as a function of $H_*$,
where the $1\sigma$ ($2\sigma$) allowed region of $n_s$ is shown by red (blue) points. 
The solution with the larger (smaller) ratio corresponds to the solution 1(2).
One can see that the ratio is indeed of order $0.1$ as expected, and it is in fact 
 a slowly decreasing function of $H_*$. This is because 
 $N_*$ increases as  $H_*$ (see Fig.~\ref{fig:e-folding}) due to \Eq{efold}, and hence a smaller $m_\phi$ is required to 
give a better fit to the observed spectral index.   For later use, let us give the relation
between  $H_*$ and the quartic coupling $\lambda$ obtained by fitting the numerical results
for $10^{-12}\GEV < H_*<10^{-8}\GEV$: 
\begin{align}
\laq{fit1}
H_* &\simeq  \left\{
	\begin{array}{ll}
	\displaystyle{2.2\, \EV \({m_\f \over 1\EV}\)^{1.007}} & ~~{\rm for~solution~1}\\
	&\\
	\displaystyle{3.1\,\EV\({m_\f \over 1\EV}\)^{1.011}}& ~~{\rm for~solution~2}
	\end{array}
	\right.,\\
\laq{fit2}
 \lambda &\simeq  \left\{
 \begin{array}{ll}
  \displaystyle{6.3 \times10^{-12}{\({m_{\f}\over 1\EV}\)}^{-0.044} }&  ~~{\rm for~solution~1}\\
&\\
  \displaystyle{9.3 \times 10^{-13}{\({m_{\f}\over 1\EV}\)}^{-0.061}}&~~{\rm for~solution~2}
 \end{array}
 \right..
\end{align}

Our results show that the inflaton mass at the potential minimum is directly related to
the Hubble parameter at the horizon exit of the CMB scales.
Using $3M_p^2H_*^2\simeq V_0$, Eq.~(\ref{lambda}) and the above two relations, 
one can express the decay constant $f$ as a function of the mass $m_\phi$,
\begin{equation}
\label{decay-mass}
f\simeq
\left\{
\begin{array}{ll}
\displaystyle{4.0\times10^7\,{\rm GeV}\, 
\left(\frac{n}{3}\right)^\frac{1}{2}
\left(\frac{m_\phi}{1\,{\rm eV}}\right)^{0.51}}
& ~~{\rm for~solution~1}\\
&\\
\displaystyle{7.7\times10^7\,{\rm GeV}\,
\left(\frac{n}{3}\right)^\frac{1}{2}
\left(\frac{m_\phi}{1\,{\rm eV}}\right)^{0.52}}
& ~~{\rm for~solution~2}
\end{array}
\right..
\end{equation}
We emphasize that the above relation is fixed by the Planck normalization of the curvature 
perturbation and the spectral index, and therefore is a rather robust prediction of our scenario. 
Note that, in contrast to the QCD axion, the decay constant becomes larger the heavier the
ALP mass is. This is because both parameters increase as the inflationary scale.

The inflaton (axion) may have couplings to the SM particles such as photons or leptons.
Such axion coupled to photons is called an ALP and has been searched
for in various experiments. As we shall see,
successful reheating after inflation can be realized by the coupling to photons. 
The ALP coupling to photons is given by
\begin{align}
\laq{int}
{\cal L} & = c_\g \frac{\a}{4 \pi} \frac{\phi}{f} F_{\mu \nu} \tilde F^{\mu \nu} 
\equiv \frac{1}{4} g_{\phi\g\g} \phi F_{\mu \nu} \tilde F^{\mu \nu}
\end{align}
where  $F^{\mu \nu}$ and $\tilde F^{\mu \nu}$ are the electromagnetic field strength and
its dual, respectively,  $c_\g$ is a model-dependent numerical factor, and $\alpha$ is the 
fine-structure constant.  In the second equality we have defined
\beq
g_{\phi\g\g} \equiv c_\g {\a \over \pi} {1\over f}.
\eeq
For instance, if there are extra fermions $\{\psi_i\}$ that transform as
$\psi_i \to e^{i \beta q_i \gamma_5/2} \psi_i$
under the shift symmetry transformation, $\phi \to \phi + \beta f$,
$c_\gamma$ is given by
\beq
c_\gamma = \sum_i q_i Q_i^2,
\eeq
where $\beta$ is the transformation parameter, $q_i$ is the charge of the $\psi_i$ under the shift symmetry, 
and $Q_i$ is its electric charge.
In most cases $c_\g$ is in the range between $10^{-2}$ and $10^2$, but it can be much 
larger or smaller if one considers an alignment or
clockwork mechanism~\cite{Kim:2004rp,Choi:2014rja,Higaki:2014pja,Harigaya:2014rga,
Choi:2015fiu,Kaplan:2015fuy,
Higaki:2015jag,Higaki:2016yqk,Giudice:2016yja, Farina:2016tgd}.\footnote{
The enhancement of the coupling to gauge bosons is simply due to the effectively large PQ charge
generated by the alignment/clockwork mechanism. See also e.g. Ref.~\cite{Sikivie:1986gq}. 
}

\begin{figure}[!t]
\begin{center}  
   \includegraphics[width=110mm]{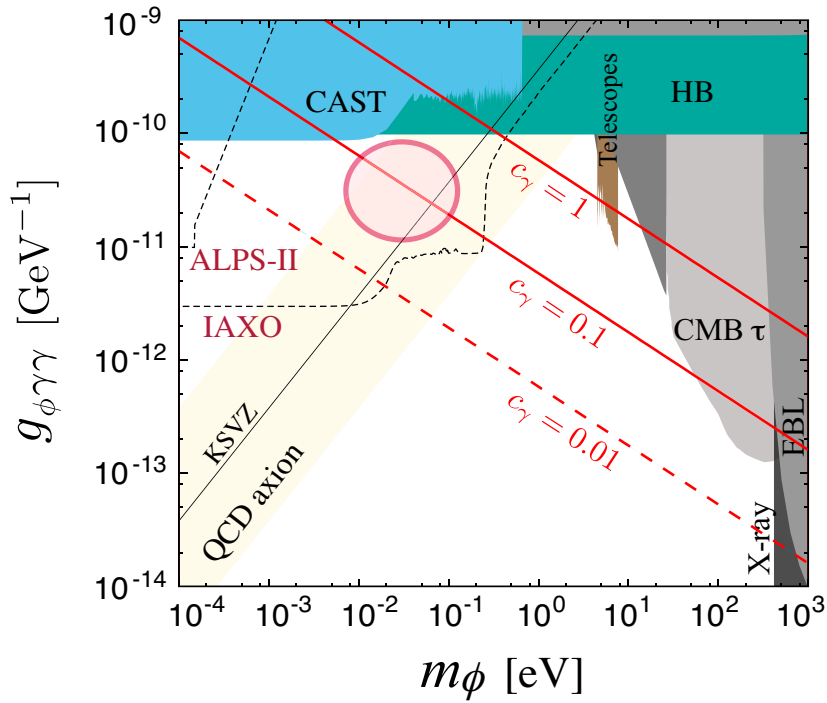}
\end{center}
\caption{
The predicted relation between the ALP (inflaton) mass $m_\phi$ and the coupling to photons $g_{\phi \g \g}$
for $c_\g = 0.01, 0.1,$ and $1$ shown as red lines. 
The dark shaded regions are excluded by various experiments and astrophysical/cosmological observaitons. 
The dashed lines represent the projected sensitivity of the next generation axion search experiments, 
ALPS-II~\cite{Bahre:2013ywa}
and IAXO~\cite{Irastorza:2011gs,Armengaud:2014gea}. For comparison, the predicted range for the QCD axion is shown as thin soild line and the light shaded (yellow) region. (Figure arranged from Ref.~\cite{Essig:2013lka}.)
The shaded circle represents a sweet spot region where the ALP miracle takes place. 
}\label{fig:infsc}
\end{figure}

In \Fig{infsc} we show the predicted relation between $g_{\phi\g\g}$ and $m_\phi$ for $c_\g = 0.01$, $0.1$,
and  $1$ as red lines. 
We have used the solution 1, but there is no signifiant change if we use the solution 2.
For comparison, we also show the predicted region of the QCD axion as a thin (black) line and the light shaded (yellow)
region, which clearly shows a different relation between the mass and coupling. 
The dark shaded regions are excluded by various experiments or 
astrophysical/cosmological observations (the exclusion limits are taken from Ref.~\cite{Essig:2013lka}). 
Note that those constraints other than CAST and horizontal branch stars assume that the ALP constitutes 
all or some fraction of dark matter. 
The dashed lines represent the sensitivity reach of the projected axion search experiments
ALPS-II~\cite{Bahre:2013ywa} and IAXO~\cite{Irastorza:2011gs,Armengaud:2014gea}. 
As one can see, $c_\g$ is constrained to be below about $\sim 3$, and  for $c_\g \gtrsim 0.1$, 
there is an allowed range for the ALP mass, $6\, {\rm meV} \lesssim m_\phi \lesssim 60$\,eV.
In addition, the parameter region will also be covered by the proposed experiments
with purely laser-based stimulated photon-photon collider~\cite{Fujii:2010is,Homma:2017cpa}.
Therefore,
our inflation model can be tested  in various ground-based  experiments. 
The circle represents a sweet spot region
where the ALP can also explain the dark matter, as we shall discuss below.

\subsection{Reheating}
Let us now discuss how the inflaton reheats the Universe. For the moment we neglect the
CP phase $\theta$ and the inflaton mass $m_\phi$. 
Such approximation is valid as long as we consider the inflaton dynamics
soon after inflation. As we shall see,  the inflaton mass becomes important when we evaluate
the abundance of the ALP dark matter.

After inflation the inflaton oscillates about the potential minimum, $\phi_{\rm min} = \pi f$.
Since the curvature of the potential vanishes at $\phi = \phi_{\rm min}$, the potential
can be well approximated by the quartic potential about $\phi_{\rm min}$,
\begin{align}
V(\phi) &\simeq  \lambda(\phi - \phi_{\rm min})^4
\end{align}
Then, the inflaton energy density scales as $R(t)^{-4}$ like radiation,
where $R(t)$ is the scale factor. 
The curvature of the potential, $V''(\phi)$, depends on the field value. 
Let us define the effective mass by $V''(\phi)$,
\beq
\label{meff}
m_{\rm eff}^2(t)\equiv V''(\phi_{\rm amp}) = 12 \lambda\,\phi_{\rm amp}^2,
\eeq
where $\phi_{\rm amp}$ denotes the oscillation amplitude about $\phi_{\rm min}$.
The effective mass, $m_{\rm eff}$, therefore decreases as $R(t)^{-1}$.

If the inflaton has a coupling to photons as Eq.~\eq{int}, the (averaged) perturbative decay rate
is given by
\beq
\laq{dr1}
\Gamma_0 (\phi \to \gamma \gamma) = \frac{\alpha^2 c_\g^2}{64 \pi^3} \frac{m_{\rm eff}^3}{f^2},
\eeq
which decreases faster than the Hubble parameter. Therefore, even if the ALP decayed at the rate given above
(without any thermal effects), it could not complete the reheating.  In fact, soon after a small fraction of the inflaton
decays into photons, the produced photons quickly form thermal plasma, and
its back reaction to the inflaton decay becomes relevant. 

Due to the finite temperature effects,
there are roughly two changes. Firstly, there appears a thermal blocking effect. 
In thermal plasma produced by the inflaton decay, photons acquire a thermal mass $m_\g^{(th)}$
of order $e T$, where $e$ is the electromagnetic  gauge coupling. 
If the thermal mass of photons exceeds the effective inflaton mass, the inflaton decay is kinematically
blocked. Thus the decay rate is modified as
\beq
\Gamma_{\rm dec} (\phi \to \gamma \gamma) = \frac{c_\g^2\alpha^2 }{64 \pi^3} 
\frac{m_{\rm eff}^3}{f^2}\sqrt{1-\left(\frac{2m_\g^{(th)}}{m_{\rm eff}}\right)^2}~~~{\rm for~~}
m_{\rm eff} \geq 2m_\g^{(th)}.
\eeq
Secondly, the inflaton can evaporate through a dissipation effect~\cite{Yokoyama:2005dv,Anisimov:2008dz,Drewes:2010pf,Mukaida:2012qn,Mukaida:2012bz}.
For instance  the inflaton dissipates its energy via a scattering process, 
$\phi + \gamma\to e^+ + e^-$. The dissipation of the QCD axion coupled to gluons was studied in 
Ref.~\cite{Moroi:2014mqa}, and it was pointed out that the dissipation rate is accompanied with a suppression
factor, $p^2/g_s^4 T^2$, when the typical momentum of the axion $p$ is smaller than $g_s^2 T$, where $g_s$ is the
strong gauge coupling. In our case, the momentum is to be replaced with the effective mass.\footnote{
\label{ft1}
Precisely speaking, even though the inflaton is initially (almost) spatially homogeneous, 
and it quickly becomes spatially inhomogeneous due to the tachyonic preheating~\cite{Felder:2000hj,Felder:2001kt}. 
The typical peak momentum is about the effective inflaton mass or less~\cite{Brax:2010ai}. 
Therefore, the tachyonic preheating increases the dissipation rate by a factor of unity, and 
does not change the argument much.}
Applying their estimate  to the ALP coupled to photons, 
we use the following dissipation rate, 
\beq
\label{Gg}
\Gamma_{{\rm dis},\gamma} = C \frac{c_\gamma^2 \alpha^2T^3}{8 \pi^2f^2} \frac{m_{\rm eff}^2}{e^4T^2}
\eeq
where $C$ is a numerical factor of ${\cal O}(1-10)$ that represents uncertainties of the
above order-of-magnitude estimate as well as the effect of tachyonic preheating.

 At temperatures higher than the electroweak scale, $T>T_{\rm EW}$, one should
consider the ALP coupling to SU(2)$_L$ and U(1)$_Y$ gauge bosons:
\begin{align}
\laq{aWWBB}
{\cal L} & = c_2 \frac{\a_2}{8 \pi} \frac{\phi}{f} W_{\mu \nu} \tilde W^{\mu \nu} 
+ c_Y \frac{\a_Y}{4 \pi} \frac{\phi}{f} B_{\mu \nu} \tilde B^{\mu \nu},
\end{align}
where $W_{\mu \nu}$ and $B_{\mu \nu}$ are the gauge field strengths of
 SU(2)$_L$ and U(1)$_Y$ gauge bosons, respectively. 
 For instance, if there are extra fermions in the fundamental representation  of
 the SU(2)$_L$, $c_2$ is given by
 \beq
 c_2 = \sum_i q_i,
 \eeq
 where $q_i$ is the charge of $\psi_i$ under the shift symmetry. 
 Similarly, $c_Y$ is given by
 \beq
 c_Y = \sum_j q_j Y_j^2,
 \eeq
 where $Y_j$ is the hypercharge of the $i$-th  chiral fermion.
 The relation between
 the coefficients is given by
 \beq
 \label{cgcwcy}
 c_\gamma = \frac{c_2}{2} + c_Y.
 \eeq
We use the following dissipation rate at temperature higher than the electroweak scale,
\beq
\Gamma_{{\rm dis,EW}} = C' \frac{c_2^2 \alpha_2^2T^3}{32\pi^2f^2} \frac{m_{\rm eff}^2}{g_2^4T^2}
+C'' \frac{c_Y^2 \alpha_Y^2T^3}{8 \pi^2f^2} \frac{m_{\rm eff}^2}{g_Y^4T^2},
\eeq
where $\alpha_2 = g_2^2/4\pi$ and $\alpha_Y = g_Y^2/4\pi$ denote the corresponding gauge couplings,
and $C'$ and $C''$ represent the uncertainties. As one can see from the relation (\ref{cgcwcy}),
if there is a mild cancellation of e.g. $10\%$ between $c_2$ and $c_Y$,  the dissipation rate 
in the early Universe can be enhanced by a factor of order $10^2$, compared to naively extrapolating
the dissipation rate (\ref{Gg}) to high temperatures.

Taking account of the above thermal effects, we have numerically solved the following
equations, 
\begin{align}
\left\{
	\begin{array}{ll}
	\displaystyle{\dot{\rho}_\phi+4H\rho_\phi=-\Gamma_{\rm tot}\rho_\phi} \\
	&\\
	\displaystyle{\dot{\rho}_r+4H\rho_r=\Gamma_{\rm tot}\rho_\phi}\label{evolution}
	\end{array}
	\right.,
\end{align}
where $\rho_\phi$ and $\rho_r$ denote the energy densities of the inflaton and radiation, respectively,
and $\Gamma_{\rm tot} = \Gamma_{\rm dec} + \Gamma_{\rm dis, \gamma}
+ \Theta(T/T_{\rm EW}) (\Gamma_{\rm dis, EW}-\Gamma_{\rm dis, \gamma})$.
If the total decay rate $\Gamma_{\rm tot}$ becomes much larger than  $H_{\rm inf}$, 
the dissipation process is very effective, and almost all the inflaton energy is immediately converted 
to the plasma i.e. instantaneous reheating takes place. 
The reheating temperature in this case is estimated as
\begin{equation}
\label{TR}
T_{R} \simeq 
\left\{
\begin{array}{ll}
\displaystyle{3.9\times10^4\,{\rm GeV}\left(\frac{g_*(T_R)}{106.75}\right)^{-1/4}\left(\frac{m_\phi}{1\,{\rm eV}}\right)^{0.50}}
& ~~{\rm for~solution~1}\\
&\\
\displaystyle{4.7\times10^4\,{\rm GeV}\left(\frac{g_*(T_R)}{106.75}\right)^{-1/4}\left(\frac{m_\phi}{1\,{\rm eV}}\right)^{0.51}}
& ~~{\rm for~solution~2}
\end{array}
\right.,
\end{equation}
where $g_*(T)$ is the effective relativistic degrees of freedom contributing to the energy density. 

We have numerically confirmed that the instant reheating indeed takes place for certain parameters. 
To parametrize the efficiency of the instant reheating due to the dissipation process, let us
define a ratio of the energy densities after the instant reheating:
\beq
\xi \equiv \left.\frac{\rho_\phi}{\rho_\phi + \rho_R}\right|_{\rm after\,instant\,reheating}.
\eeq
For instance, we obtained $\xi  \simeq 0.02$ for $c_2 =3 $, $c_Y =-5/4$,
$g_{\phi \g \g} \simeq 7 \times 10^{-11}$\, GeV$^{-1}$, $C = C' =C'' = 4$. 
Thus, the instant reheating due to the dissipation is indeed possible for those parameters
satisfying the existing limits on the ALP-photon coupling. 
While it is fair to say that the above result is obtained based on a rather crude approximation and so 
it still contains a large uncertainty\footnote{\label{f7}
Note that the relation between $m_\phi$ and $f$ depends on $n$ and that it is also slightly modified if
we vary the relative height of the two cosine terms in the inflaton potential~\cite{Czerny:2014xja}. 
We find that, by varying $\kappa$,  $m_\phi$ can increase at least by several tens percent for a fixed $f$,
which enhances the ratio $\Gamma_{\rm dis}/H_{\rm inf}$ by a factor of unity.},
it is probably difficult to realize $\xi \lesssim 0.01$ for $g_{\phi \g \g} \lesssim 10^{-11}$\,GeV$^{-1}$.
 A more precise calculation is certainly necessary to delineate the parameter region
 where the instant reheating takes place. 
  There are a couple of ways to enlarge the parameter region for successful reheating.
One is to consider multiple axions, which will be studied in the next section. Another is
to introduce couplings to leptons.\footnote{
In particular, a coupling to $\tau$ is important because the dissipation rate is proportional
to the lepton mass squared. The ALP coupling to photons and electrons can be suppressed 
if one considers an anomaly-free charge assignment, e.g. $q_\tau = -q_\mu$,
which allows a smaller decay constant $f$ evading experimental/observational limits~\cite{Nakayama:2014cza}.
} We leave detailed studies on these possibilities for future work.

Lastly let us mention that the ALP inflaton is (almost) thermalized during the dissipation process,
and it gets soon decoupled (cf. Ref.~\cite{Salvio:2013iaa}). The abundance of thermalized ALPs is expressed in terms of 
the effective number of extra neutrino species,
\beq
\label{Neff}
\Delta N_{\rm eff} \simeq 0.027 \left(\frac{106.75}{g_{*s}(T_R)}\right)^\frac{4}{3},
\eeq
where $g_{*s}(T)$ is the effective relativistic degrees of freedom contributing to the entropy density.
The temperature of the thermalized ALPs is given by
\beq
T_{\rm \phi} \simeq 0.33 \left(\frac{106.75}{g_{*s}(T_R)}\right)^\frac{1}{3} T_\gamma,
\eeq
where $T_\gamma$ is the photon temperature after the electron-positron annihilation.
As we shall see soon,  the inflaton mass $m_\phi$ should be of order $0.01$\,eV for the remaining condensate 
to account for cold dark matter. Then, thermalized components become non-relativistic soon after recombination, 
and they behave as hot dark matter. See e.g. Ref.~\cite{Bridle:2016isd} for the recent bound on the abundance and effective mass
of such hot dark matter component. 
The prediction of such hot dark matter with the above abundance is a rather robust prediction of our scenario, and it 
can be tested by future CMB and BAO measurements~\cite{Kogut:2011xw,Abazajian:2016yjj,Baumann:2017lmt}.

\subsection{ALP dark matter}
Next we shall discuss requirements for the remaining ALP to account for dark matter\footnote{
The axion dark matter in a flat-bottomed potential was studied by Takeshi Kobayashi and
two of the present authors (RD and FT) in Ref.~\cite{Daido:2016tsj}, where the abundance
and isocurvature perturbations are shown to be suppressed compared to the potential
with a single cosine term. }  See e.g. Refs.~\cite{Kim:2008hd,Bae:2008ue,Wantz:2009it,Ringwald:2012hr,Kawasaki:2013ae,Marsh:2015xka} for recent reviews on axion dark matter.
So far we have neglected the ALP mass; the ALP inflaton oscillates in a quartic potential and
its energy density decreases like radiation.
 However, the ALP mass
becomes relevant when the oscillation amplitude becomes sufficiently small.\footnote{
The ALP condensate actually has a nonzero momentum due to the tachyonic preheating (cf. footnote \ref{ft1}).
The momentum becomes negligible almost at the same time since the peak frequency is comparable
to or less than the effective mass. } 
 The transition from the quartic potential to the quadratic one takes place when the 
amplitude becomes comparable to $\phi_c = m_\f/\sqrt{2 \lambda}$.  Afterwards the ALP energy
density behaves like non-relativistic matter and contributes to cold dark matter.

The ALP abundance  and  the transition temperature $T_c$ are related by
\beq
\frac{\rho_\phi}{s} = \frac{3 T_c}{4} \frac{\xi}{1-\xi} 
\left(\frac{g_{*s}(T_R)}{g_{*s}(T_c)}\right)^\frac{1}{3}.
\eeq
Assuming that the ALP condensate
accounts for the observed dark matter density, we can express $T_c$ in terms of $\xi$ as
\beq
T_c \simeq  0.6\, \xi^{-1} \left(\frac{g_{*s}(T_R)}{g_{*s}(T_c)}\right)^\frac{1}{3} 
\left(\frac{\Omega_{\rm \phi}h^2}{0.12}\right) 
\,{\rm eV},
\eeq
which implies that the dark matter is formed between the matter-radiation equality and big bang
nucleosynthesis unless $\xi$ is extremely small.  
Here $\Omega_\phi$ is the density parameter of the ALP dark matter, and the observed 
dark matter abundance is $\Omega_{\rm DM}h^2 \simeq 0.12$~\cite{Ade:2015lrj}.
Since the ALP condensate behaves like dark radiation at $T > T_c$,  it
suppresses the small-scale matter power spectrum like late-forming dark matter 
or warm dark matter. In order to
be consistent with the SDSS and/or Lyman-$\alpha$ data, 
the transition should take place no later
than the redshift $z_c = {\cal O}(10^5)$~\cite{Sarkar:2014bca}. Then the ratio $\xi$ is
bounded above:
\beq
\label{xi}
\xi \lesssim 0.02 
\left(\frac{g_{*s}(T_R)}{106.75}\right)^\frac{1}{3}
\left(\frac{3.909}{g_{*s}(T_c)}\right)^\frac{1}{3} \left(\frac{\Omega_{\rm \phi}h^2}{0.12}\right)
\left(\frac{5 \times 10^5}{1+z_c}\right).
\eeq
Therefore, the small-scale matter power spectrum sets the lower bound on the
dissipation rate, i.e. on the ALP-photon coupling up to the above mentioned 
uncertainties. Note that here we have not used any information from the 
requirement that the ALP should drive successful inflation.

Now let us express the abundance of the ALP condensate in another way and 
derive a bound on the ALP mass.
In the following we approximate $\xi \ll 1$ to satisfy the above bound. 
After the instant reheating, the oscillation amplitude of the ALP condensate
becomes smaller by a factor of $\sim \xi^\frac{1}{4}$. 
The oscillation amplitude is written as $\phi_{\rm amp} = \xi^\frac{1}{4} x f$, where $x$ is a
numerical factor of order unity.
After the instant reheating, the energy density of the ALP condensate
decreases as $R(t)^{-4}$ until $\phi_{\rm amp}$  becomes smaller than $\phi_c$.
Then the abundance is given by
\begin{align}
\frac{\rho_\phi}{s} &\simeq \frac{3}{4} \xi^\frac{3}{4} \frac{m_\phi T_R}{\sqrt{2 \lambda} f x}.
\end{align}
Equating this to the dark matter density, we obtain
\begin{align}
\label{mbelow}
m_\phi &\simeq 0.07\, x^{-1} \left(\frac{\xi}{0.01}\right)^{-\frac{3}{4}}  \left(\frac{\Omega_{\rm \phi}h^2}{0.12}\right) {\rm eV},\\
&\gtrsim 0.04 \, x^{-1}
\left(\frac{106.75}{g_{*s}(T_R)}\right)^\frac{1}{4}
\left(\frac{g_{*s}(T_c)}{3.909}\right)^\frac{1}{4}  \left(\frac{\Omega_{\rm \phi}h^2}{0.12}\right)^\frac{1}{4} 
 \left(\frac{1+z_c}{5 \times 10^5}\right)^\frac{3}{4}  {\rm \,eV},
 \label{mbelow2}
\end{align}
where we have used  Eq.~(\ref{TR}) and the solution 1 of Eq.~(\ref{decay-mass}) in the
second line. For the solution 2, the numerical coefficients 0.07 and 0.04 should be replaced with 0.04 and 0.03, respectively.
 Thus, the ALP mass should be heavier than roughly $\sim 0.01$\,eV
to explain all the dark matter and inflation. It is however non-trivial if such ALP can efficiently dissipate
its energy to satisfy the bound (\ref{xi}), because $g_{\phi \gamma \gamma}$ tends to decrease as $m_\phi$ increases. 

\subsection{The ALP miracle}
Here let us summarize what we have seen in this section. First of all, in the axion
hilltop inflation (\ref{DIV}), the relation between $m_\phi$ and $f$ is fixed by the
Planck normalization of the curvature perturbation and the spectral index
as Eq.~(\ref{decay-mass}). Then we have introduced a coupling to photons and shown
that the instant reheating after inflation is possible for certain parameters
taking account of uncertainties. This basically sets a lower bound 
$g_{\phi \gamma \gamma} \gtrsim 10^{-11}$\,GeV$^{-1}$,
which is equivalent to a certain upper bound on $m_\phi$ (up to a model-dependent factor $c_\gamma$) 
if we use the relation (\ref{decay-mass}).
On the other hand, the ALP mass is bounded below as (\ref{mbelow2}) to explain
the observed dark matter density and satisfy the current observations of the small-scale
matter power spectrum. It is therefore quite non-trivial if there is an allowed region satisfying
the upper and lower bounds on $m_\phi$. In fact, there {\it is} a parameter region
satisfying all the requirements without running afoul of the experimental/observational constraints:
\begin{align}
\label{ssm}
m_\phi &= {\cal O}(0.01) {\rm\,eV},\\
\label{ssg}
g_{\phi \gamma \gamma} & = {\cal O}(10^{-11}) {\rm GeV}^{-1}.
\end{align}
Note that, for a heavier ALP mass, the parameter $\xi$ must be lower, which requires a 
larger value of $g_{\phi \gamma \gamma}$ in a tension with the CAST or astrophysical bounds. 
This sweet spot of our scenario is shown by a circle in Fig.~\ref{fig:infsc}. We call this coincidence
as the ALP miracle.  Furthermore, this sweet spot is within the sensitivity reach of the IAXO 
experiment~\cite{Irastorza:2011gs,Armengaud:2014gea} and the proposed purely laser-based 
stimulated photon-photon collider \cite{Fujii:2010is,Homma:2017cpa}.\footnote{
{A purely laser-based stimulated photon-photon collider concept
was advocated in Ref.~\cite{Fujii:2010is}, and searches  based on the concept
were demonstrated for the mass below $0.1$ eV~\cite{Homma:2014rja,Hasebe:2015jxa}.
The extension of the same method up to the $0.1$~eV - $10$~keV mass range was also
proposed in Ref.~\cite{Homma:2017cpa}. 
This experimental approach will also provide opportunities to test our model. }}

We also note that there is a hint to an ALP couple to photons in a study of the so called 
$R$-parameter, the ratio of the number of horizontal branch stars to red giants in globular 
clusters~\cite{Ayala:2014pea,Straniero:2015nvc}. The latest analysis shows
$g_{\phi \gamma \gamma} = (0.29 \pm 0.18) \times 10^{-10} {\rm\,GeV}^{-1}$~\cite{Straniero:2015nvc},
which overlaps with the sweet spot of the ALP miracle. 
Lastly we emphasize again that this scenario
predicts a hot dark matter composed of thermalized ALPs, whose abundance is given by
$\Delta N_{\rm eff} \sim 0.03$ (cf. Eq.~(\ref{Neff})).

\section{Extension to multi-axion fields}
\label{sec3}

In the previous section, we have seen that, in the axion hilltop inflation with $n$ being an odd number,
the inflaton is so long-lived that it still remains until present and contributes to  dark matter.
In particular, if the axion is coupled to photons, both the observed curvature perturbation and dark matter 
abundance can be simultaneously explained for the ALP mass and coupling to photons given by 
(\ref{ssm}) and (\ref{ssg}).
 We note however that the dissipation rate used in our calculation is still a crude
order-of-magnitude estimate, and a more precise calculation is necessary to narrow 
the whole parameter space.

In this section we consider an extension of the previous model to multiple axion fields. We show that
the dissipation rate of the inflaton can be enhanced by an order of magnitude or more,
which enlarges the parameter region where both inflation and dark matter can be explained
while satisfying the current constraints on the coupling to photons. 

As an concrete example, we consider an axion hilltop inflation model with two axions, where
the inflaton dynamics can be similarly realized by identifying one combination of the two
axions with the inflaton. 
There are roughly two possible cases. One is that the axion inflation
ends like hybrid inflation~\cite{Copeland:1994vg,Dvali:1994ms,Linde:1997sj}, 
and the heavy axion is mainly generated after inflation, which
decays and dissipates into plasma through an enhanced coupling to photons. The enhancement is 
due to the alignment mechanism~\cite{Kim:2004rp}. For a mixing of order unity, the
dissipation rate can be enhanced by an order of magnitude, which will enlarge the parameter
region of the ALP miracle. Further enhancement is also possible in a model with multiple 
 axions with the alignment/clockwork structure. The other is that, although the inflationary
direction during and after inflation mainly coincides with the light axion, 
the heavier axion is resonantly produced when the two oscillation frequencies become comparable
to each other. The resonance exhibits a chaotic behavior, and we call this phenomenon as the chaotic
conversion.  The heavy axion produced by the
chaotic conversion decays/dissipates into radiation. In either case,
if the heavy axion dominates the Universe and produces entropy by its decay, 
the light axion abundance can be reduced, in which case the observed dark matter
abundance can be explained by the light axion of mass heavier than $0.01$\,eV.

Another interesting aspect of the model with multiple axions is that some of the heavier axions
may be searched for in the beam dump experiment like SHiP~\cite{Alekhin:2015byh,Dobrich:2015jyk}. In this case,  the decay of the heavy
axion responsible for the reheating (i.e. the big bang!) can be observed at the ground-based experiments.

\subsection{Axion hilltop inflation with two axions}
\begin{figure}[t!]
\begin{center}  
   \includegraphics[width=70mm]{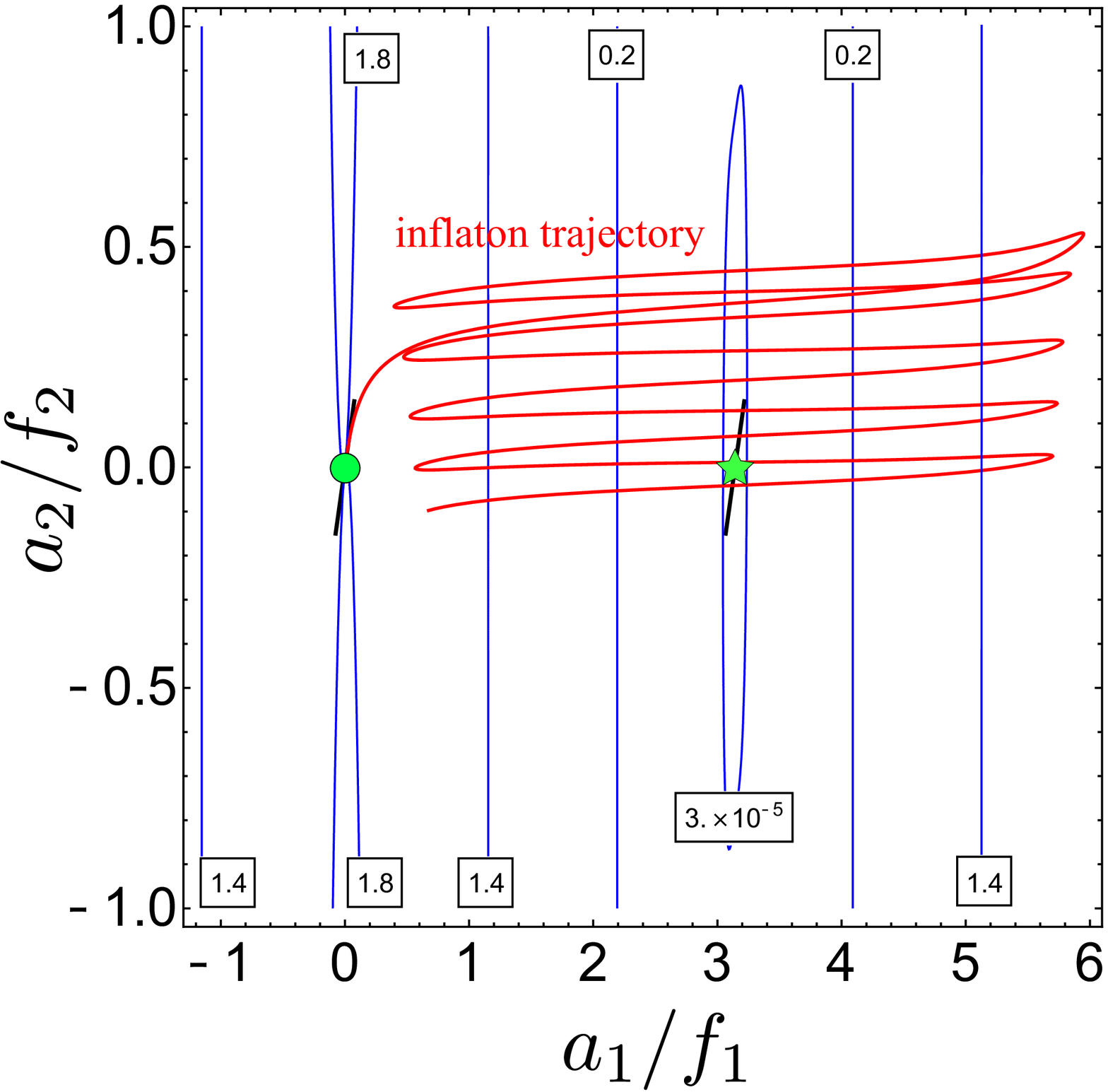}
   \includegraphics[width=78mm]{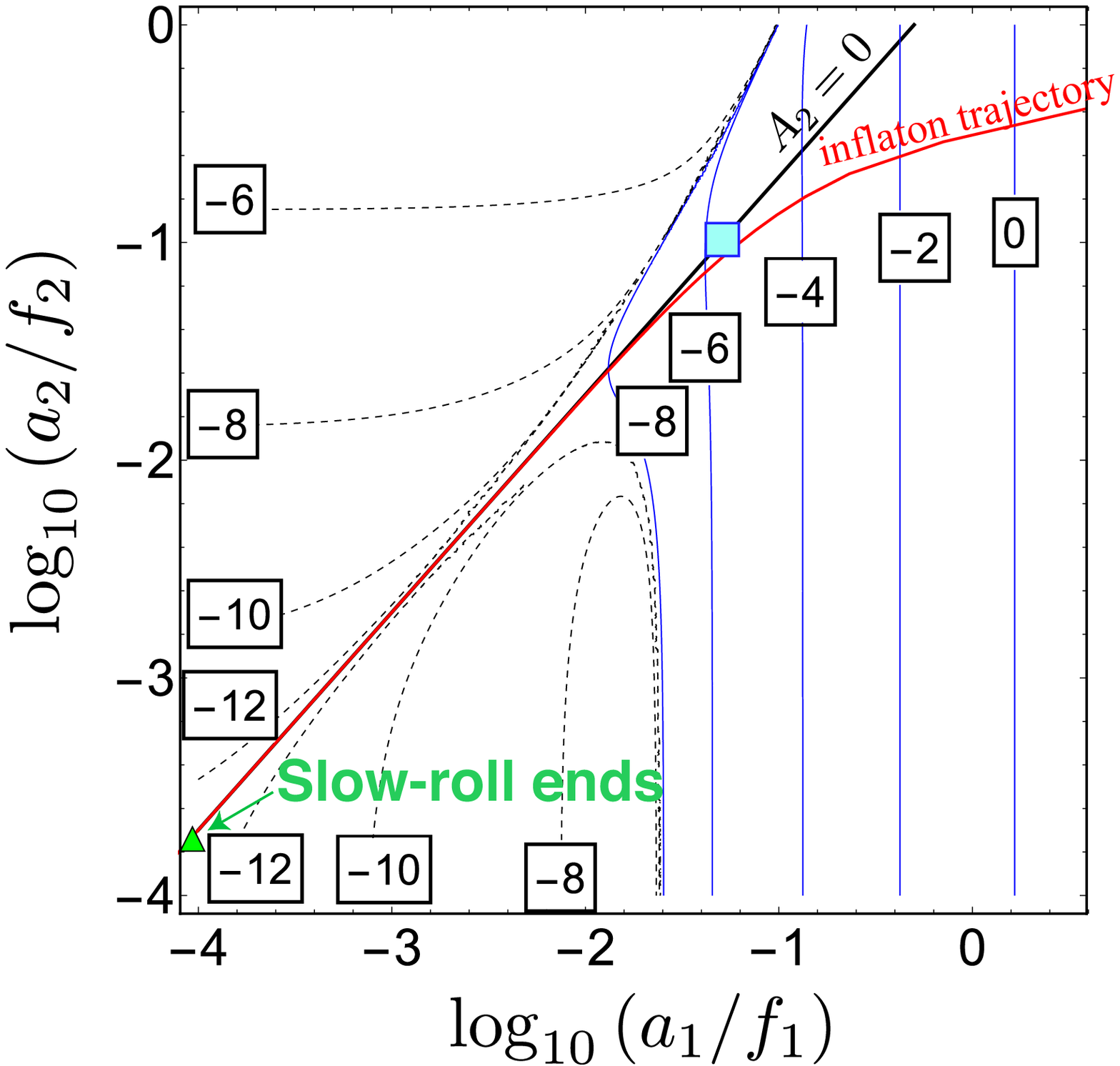}
\end{center}
\caption{The contours of the inflaton potential, 
$V(a_1,a_2)/ \Lambda_1^4$ (left) and $\log_{10}{\frac{\ab{(V(a_1,a_2)-V(0,0))}}{\GEV^4}}$ (right),
 with  $\Lambda_1=10^{12}\, \GEV,f_1=10^{15}\,\GEV,f_2=10^{14}\,\GEV, \Lambda_2=10^{11}\,\GEV,k=-2 \AND \h=0$. 
 The red solid line represents the trajectory of the  inflaton. 
The green disc and star represent the saddle point and minimum, respectively, and the short line segments (black) represent $A_H = 0$, i.e., the direction the lightest axion, $A_L$.
In the right panel, the slow-roll ends at the green triangle where $|\eta(A_L)|$ becomes unity, 
and the curvatures along $A_L$ and $A_H$ become equal at the (light blue) square.
Afterwards the inflaton evolves toward the potential minimum, and it is $a_1$ that is mainly produced. 
}
\label{pt}
\end{figure}

\lac{rc}
Now let us consider an axion hilltop inflation with two axions $a_1$ and $a_2$ 
with the following potential,
\begin{align}
\laq{intV}
V(a_1,a_2) 
&= 
\Lambda_1^4\(\cos\({a_1 \over f_1 } + \theta \)-{1 \over n^2}\cos\({n a_1\over f_1 }\)\)
-\Lambda_2^4 \cos{\( k {a_1 \over f_1}+ {a_2 \over f_2 }\)} + {\rm const.},
\end{align}
where the first term is same as the potential for the axion hilltop inflation (\ref{DIV})
with $\kappa = 1$, the second term represents 
a mixing between $a_1$ and $a_2$,
$f_1$ and $f_2$ are the decay constants of $a_1$ and $a_2$, respectively, and $k$ is a rational number.
For later use we define $$b \equiv {f_2k \over f_1},$$ which represents a mixing parameter between $a_1$ and $a_2$.
When $b=0$, there is no mixing between the two axions.
In the following we assume that $k$ is an even integer for simplicity, in which case $a_2$ stays
in the vicinity of the same minimum during and soon after inflation (See Fig.~\ref{pt}).\footnote{ 
If this is not the case, $a_2$ starts to oscillate about the adjacent minimum after inflation,
but its initial oscillation energy is subdominant compared to the initial inflaton energy.}

We assume that a combination of the two axions, $a_2 + b a_1$, acquires a heavy
mass from the second term in \Eq{intV} so that it is stabilized at the origin during inflation. 
The other orthogonal combination, $a_1 - b a_2$,  remains light and becomes the inflaton. 
Let us express these
combinations as
\begin{align}
\label{A1}
A_L & \equiv \frac{1}{\sqrt{1+b^2}} \left(a_1 - b a_2 \right),\\
A_H & \equiv \frac{1}{\sqrt{1+b^2}} \left(a_2 + b a_1 \right),
\label{A2}
\end{align}
which correspond to the mass eigenstates at the potential extrema, $(a_1, a_2) = (0,0)$ and $(\pi f_1,0)$
when $\theta = 0$. In fact, the effect of $\theta \ne 0$ on the mass eigenstates is negligible.
 We also define the decay  constants of $A_L$ and $A_H$ as
\begin{align}
F_L & \equiv\sqrt{1+b^2} f_1, \\
F_H & \equiv \frac{1}{\sqrt{1+b^2}} f_2.
\end{align}
The heavy axion $A_H$ acquires a mass,
\beq
\laq{m2}
M_H \equiv {\Lambda_2^2 \over F_H}= \sqrt{ b^2 +1}{\Lambda_2^2 \over f_2},
\eeq
and we assume that it is heavier than the Hubble parameter during inflation,
$M_H \gg H_{\rm inf}$, so that the heavier axion is stabilized at the origin,
\beq
\laq{inicon}
\langle A_H \rangle_{\rm inf} \simeq 0.
\eeq
Then, $A_H$ can be integrated out  during inflation, and
 the potential for $a_1$ in \Eq{intV} is equivalent to the inflaton potential (\ref{DIV})
with the following identification:
\beq
\laq{change}
A_L\->  \f \,
~~
F_L \-> f,
~{\rm and}~
\Lambda_1 \-> \Lambda.
\eeq  
Therefore, the results of the axion hilltop inflation in the previous section can be applied to the present case
with the two axions. In particular, the relation between the mass and decay constant (\ref{decay-mass})
still holds in this case. When $\theta$  takes a nonzero value of order $(F_L/M_{pl})^3$, the
above defined $A_L$ and $A_H$ still provide a good approximation to the mass eigenstates, 
as long as the dynamics during inflation and the axion search experiments
are concerned. 

The inflation ends when the curvature of the potential along $A_L$ becomes comparable to
$H_{\rm inf}$. The dynamics of the two axions become slightly involved afterwards.\footnote{
The natural inflation with two axions was studied in Ref.~\cite{Peloso:2015dsa} where
the inflaton path after inflation is similar to the present case, although they focused on large-field inflation.
} As the inflaton $A_L$ further rolls down, the (negative) curvature of the potential grows, and
at a certain point, it becomes comparable to the heavier axion mass, $M_H$. After that, the heavier
axion $A_H$ cannot be integrated out anymore, and both $a_1$ and $a_2$ start to oscillate about the 
potential minimum, $(a_1, a_2) = (\pi f_1, 0)$.
Since the inflaton potential (the first term in Eq.~\eq{intV}) is given in terms of $a_1$ only, 
it is $a_1$ that acquires the most of the energy after inflation.
Thus, the Universe is dominated by the coherent oscillations of $a_1$ after inflation.\footnote{
If we modify the inflaton potential, it is possible for $a_2$ acquires a larger oscillation energy just
after inflation. This can be realized e.g. by replacing the variable of the second cosine term
in the first parenthesis of \Eq{intV} with $n a_1/f_1 + k' A_H/F_H$. The inflaton dynamics during
inflation remains unchanged because $A_H$ is stabilized at the origin. However, after inflation,
$a_2$ can oscillate, acquiring some fraction of the inflaton energy. To be precise, in this modified
model,  the inflaton mass does not vanish at the potential minimum and maximum even 
when $\theta=0$, but it can be negligibly small if $\Lambda_2 \gg \Lambda_1$.}

In Fig.~\ref{pt} we show the contours of the potential, $V(a_1,a_2)/ \Lambda_1^4$ (left) 
and $\log_{10}\left|{\frac{V(a_1,a_2)-V(0,0))}{\GEV^4}}\right|$ (right),  with  
$\Lambda_1=10^{12}\, \GEV,f_1=10^{15}\,\GEV,f_2=10^{14}\,\GEV, \Lambda_2=10^{11}\,\GEV,k=-2 \AND \h=0$. 
 In the right panel, the blue solid (black dashed) lines mean
negative (positive) values of $V(a_1,a_2)-V(0,0)$.  One can see that the inflationary path is along the 
valley of $A_H = 0$, which is represented by the black solid line.
The inflaton trajectory is shown by the red line.
The slow-roll inflation ends at the green triangle in the right panel, and 
the inflaton continues to roll down for a while. Then, the curvature of the potential
along $A_L$ becomes comparable to $M_H$ around the (light blue) square. Afterwards,
both  $a_1$ and $a_2$ starts to evolve, and the inflaton oscillates about the minimum mainly 
along the direction of $a_1$.
(In the figure we have chosen parameters so that the oscillations along $a_2$ are visible.)

Let us note here that the inflaton dynamics after inflation can be broadly classified
into the two cases, (i) $|b| > 1$ and (ii) $|b| < 1$.
If $|b| > 1$, the light axion, $A_L$, which plays the role of the inflaton
and dark matter, mainly consists of $a_2$, while the heavy axion, $A_H$,
mainly consists of $a_1$ (cf. Eqs.~(\ref{A1}) and (\ref{A2})). In this case,
the inflaton dynamics resembles the hybrid inflation:  inflation is driven
by the light axion $A_L$, and after inflation,  the Universe is dominated
by the oscillation energy of $a_1$, which is mostly the heavy axion $A_H$.
(Note that the mass eigenstates $A_L$ and $A_H$ are defined only at the
potential maximum and minimum, and they are not necessarily a
good description when the axion oscillates with a large amplitude.)
On the other hand,  if $|b| < 1$,  $A_L$ and $A_H$ mainly consist of $a_1$ and
$a_2$, respectively, and the heavy axion does not play any important role during
and soon after inflation. As we shall see next, however, the heavy axion can be resonantly 
produced when the two oscillation frequencies of $a_1$ and $a_2$ become comparable
to each other unless $|b| \ll 1$.

To summarize, the axion hilltop inflation can be similarly realized in a model with two axions, and in particular,
the inflaton dynamics and therefore the predicted properties of the curvature perturbation are the same
as before. On the other hand, the dynamics after inflation is more involved, and the reheating process
can be modified accordingly, which will be studied in the next subsection.

\subsection{Coupling to photons and reheating} 
\lac{ctp}
Now let us study couplings of the axions to photons and discuss
how the inflaton reheats the Universe through the interaction. 
To be concrete, we assume the following coupling to photons,
\begin{align}
\frac{\alpha}{4\pi} \left(c_{\gamma 1} \frac{a_1}{f_1} + 
c_{\gamma 2} \frac{a_2}{f_2} \right) F_{\mu \nu} \tilde{F}^{\mu \nu},
\end{align}
where $c_{\gamma 1}$ and $c_{\gamma 2}$ denote the coupling
constants and depend on the number and charge assignment of 
those particles coupled to the axions. The extension to the coupling to
the weak gauge bosons is straightforward. 

As we have seen above, the inflaton dynamics after inflation can be broadly
classified into two cases depending on whether $|b|$ is greater than unity or not. 
Below  we consider the following two cases separately,
(i) $|b| > 1$ with $c_{\gamma 1} \ne 0$ and $c_{\gamma 2} = 0$, and 
(ii) $|b| < 1$ with $c_{\gamma 1} = 0$ and $c_{\gamma 2} \ne 0$.

\subsubsection{Case of $|b| > 1$ with $c_{\gamma 1} \ne 0$ and $c_{\gamma 2} = 0$} 
\lac{hybrid}
What is different from the single axion model is that, while the inflation is driven by $A_L$
(made mostly of $a_2$), the Universe after inflation is dominated by the coherent oscillations 
$a_1$. Since $a_1$ is more strongly coupled to photons than $A_L$, 
one can enhance the dissipation rate in comparison to the single axion case. 
To see this, let us rewrite the
axion photon coupling in terms of $A_L$ and $A_H$:
\begin{align}
\frac{\alpha}{4\pi} c_{\gamma 1} \frac{a_1}{f_1} F_{\mu \nu} \tilde{F}^{\mu \nu}
= \frac{\alpha}{4\pi} c_{\gamma 1} \left(  \frac{A_L}{F_L} 
+ \frac{b}{\sqrt{1+b^2}} \frac{A_H}{f_1}
\right)
F_{\mu \nu} \tilde{F}^{\mu \nu}
\end{align}
Thus, the coupling of $a_1$ to photons is stronger by a factor of $\sqrt{1+b^2}$
than $A_L$, due to the alignment mechanism~\cite{Kim:2004rp}.
(Alternatively, one can say that the coupling of $A_L$ is suppressed.)
The dissipation rate can be enhanced by a factor of $(1+b^2)$, and so, $
|b| \sim 3$ enhances the rate by an order of magnitude.
This makes it easier for $a_1$ to dissipate into plasma, while satisfying the
current limits on the coupling of $A_L$ to photons.  After the dissipation, some fraction of the
coherent oscillations of $a_1$ may remain, from which both $A_L$ and $A_H$ are produced.
The production of $A_L$ and $A_H$ may proceed resonantly (see below)\footnote{Here we assume 
that the finite temperature effect does not affect the potential \eq{intV}, especially for the second term 
which in general satisfies $\Lambda_2<\Lambda_1 \sim T_R  $ to guarantee the trajectory after the 
inflation along the $a_1$ axis. This may be due to that $\Lambda_2$ is an explicit breaking term for
 the shift symmetry of $A_H$ rather than a dynamical one. 
 Otherwise, the axion dynamics will become more involved. 
 }, in which case their number densities become comparable to each other. 
 The heavy axion $A_H$ will eventually
decay into photons, and the light axion $A_L$ remains as dark matter. Depending on the abundance
of $A_H$ and when it decays, the allowed mass range of $A_L$ will be modified. For instance,
if $|b| \sim 10$, the $a_1$ will evaporate more efficiently than the case with a single ALP for a fixed $F_L = f$. 
The abundance of $A_H$ is determined by $\xi$, which measures the abundance of the remnant $a_1$.
For $\xi \sim 0.01$, $M_H \sim 1$\,GeV, and $f_1 \sim 10^6$\,GeV, $A_H$ dominates the Universe
and its decay temperature  is about $100$\,MeV. Then, the entropy dilution factor is 
 of order $100$, and the mass of $A_L$ should be heavier by the same amount
 compared to the mass in the ALP miracle region, i.e., 
 $m_\phi =  {\cal O}(1)$\,eV,  to account for the dark matter abundance.
Note that the dark matter abundance remains
almost the same by varying $\xi$ over some range, because the production from the $a_1$ remnant  
and the dilution due to the 
decay of $A_H$ are canceled. In fact, assuming that $A_H$ dominates the Universe and 
that the number densities of $A_L$ and $A_H$
are comparable to each other due to the chaotic conversion, one can estimate the abundance of $A_L$ as
\begin{align}
\frac{\rho_{A_L}}{s} &= \frac{3 T_{\rm dec}}{4 M_H} m_\phi \frac{n_{A_L}}{n_{A_H}} \\
&\sim 10^{-10}{\rm\,GeV} \left(\frac{T_{\rm dec}}{100{\rm \,MeV}}\right)
\left(\frac{M_H}{1{\rm \,GeV}}\right)^{-1} 
\left(\frac{m_\phi}{1{\rm \,eV}}\right),
\end{align}
where $T_{\rm dec}$ is the decay temperature of $A_H$ and $n_{A_L}$ and $n_{A_H}$
are the number densities of $A_L$ and $A_H$, respectively. Interestingly, for $m_\phi ={\cal O}( 1)$\,eV and
$g_{\phi \gamma \gamma} = {\cal O}(10^{-11}) {\rm \,GeV}^{-1}$, the $A_L$ becomes
decaying dark matter which may be probed by observations (see Fig.~\ref{fig:infsc}).

If $|b| \gg 1$, $a_1$ likely evaporates completely soon after inflation 
through the coupling to photons. On the other hand, a small amount of $a_2$ will remain,
which mostly becomes $A_L$ at a later time. 
More specifically, 
both $a_1$ and $a_2$ start to oscillate about the minimum 
 when the curvature of the potential along $A_H = 0$ becomes comparable to $M_H$.
This takes place at 
\begin{align}
a_1^c & \simeq \frac{\Lambda_2^2}{\sqrt{12 \lambda} f_2},\\
a_2^c & \simeq \frac{-b \Lambda_2^2}{\sqrt{12 \lambda} f_2},
\end{align}
where $\lambda$ is given in Eqs.~(\ref{pninni}) and \eq{fit2}.
After that, most of the energy goes into the oscillation energy of $a_1$, while $a_2$ also oscillates about
the origin with a smaller frequency. In particular, if $|b| \gg 1$, most of the initial oscillation energy
of $a_2$ turns into that of $A_L$. Also, if $f_1 \ll f_2$,  $a_1$ is likely to decay and evaporate into
plasma soon after inflation, i.e.,  the instant reheating takes place.
We have both numerically and analytically checked that the resultant $\xi$ parameter tends to be so small 
 that thus produced $A_L$ cannot account for dark matter. In particular, this is considered to be the case when the heavy ALP, $A_H$,
is in the sensitivity reach of SHiP. One needs to consider other production mechanism of $A_L$ or a possibility to suppress
the evaporation rate of $a_1$. For instance, if the $\Lambda_2$ is temperature dependent and the potential for $a_2$ vanishes
when the temperature exceeds $\Lambda_2$, the oscillation of $A_L$ could be delayed until the potential appears again at a later time, 
which may enhance the abundance of $A_L$.  We leave a detailed study of such scenario for future work. 

As an extreme case, one can also consider a set-up with multiple axions based on 
a clockwork structure~\cite{Kim:2004rp,Choi:2014rja,Higaki:2014pja,Harigaya:2014rga,Choi:2015fiu,Kaplan:2015fuy, Higaki:2016yqk, Giudice:2016yja, Farina:2016tgd}, e.g., 
\begin{align}
V_{\rm inf}(a_1) -  \sum_{i=1}^{n-1} \Lambda^4 \cos\left(k \frac{a_i}{f_i} + \frac{a_{i+1}}{f_{i+1}}\right)
+ \frac{\alpha}{4\pi} c_{\gamma \ell} \frac{a_\ell }{f_\ell} F_{\mu \nu}\tilde{F}^{\mu \nu},
\end{align}
where $k (>1)$ is a numerical factor, which determines the ratio of the adjacent gears.
For simplicity we assume that all the decay constants are of the same order, $f_i \sim f$. 
During inflation, a certain combination of the axions remain light and drives the inflation.
Interestingly, the fraction of $a_1$ in the lightest axion (inflaton or dark matter)
can be suppressed by a factor of $\sim k^{n-1}$. On the other hand,
after inflation, the Universe is dominated by the coherent oscillations of $a_1$.
Now let us choose $\ell = 1$. Then, $a_1$ is coupled to photons with a decay constant $f$,
which may be strong enough for $a_1$ to dissipate into plasma and reheat the Universe. 
On the other hand, the lightest axion responsible for the dark matter is coupled to photons
with an enhanced decay constant, $F \sim k^{n-1} f$. For a sufficiently large $n$, one can
realize a hierarchy between $f$ and $F$. If we take $f \sim 10^3$\,GeV and $F \sim 10^{8}$\,GeV,
one or more of the axions may be within the reach of the beam dump experiments like SHiP.
The strength of the couplings to photons can be adjusted by changing $\ell$. For a certain choice
of $\ell$, it will be possible for the lightest axion explains both inflaton and dark matter, while
some of the heavier axions can be searched for in the SHiP experiment~\cite{Alekhin:2015byh,Dobrich:2015jyk}.

\subsubsection{Case of $|b| < 1$ with $c_{\gamma 1} = 0$ and $c_{\gamma 2} \ne 0$} 
\lac{RC}
Now let us consider the other case in which the inflaton and dark matter are explained by $A_L$ 
mainly composed of $a_1$, while the heavy axion $A_H$ is more or less decoupled
during and soon after inflation. In this case, 
the Universe is dominated by the coherent oscillations of $a_1$ for a while after inflation. However, as we shall
see below, the heavy axion is excited by the resonant process, and the initial oscillation 
energy of $a_1$ is distributed to the heavy and light axions.

In order to study the resonant process, we focus on the axion dynamics around the potential minimum, 
$(a_1, a_2) = (\pi f_1, 0)$,  and expand the potential as
\begin{align}
\laq{DMV}
V(\tilde{a}_1,a_2) &= \lambda' \tl{a}_1 ^4  +{\Lambda_2^4 \over 2 f_2^2}\({ {b \tl{a}_1 +{a_2} }}\)^2 + \cdots,\\
\laq{DMV2}
& = \frac{\lambda'}{(1+b^2)^2} \left(\tilde{A}_L+b \tilde{A}_H\right)^4  + \frac{M_H^2}{2} \tilde{A}_H^2 + \cdots,
\end{align}
where $\lambda'$ coincides with $\lambda$ in the limit of $b \to 0$, but is slightly different from $\lambda$ for $0<|b|\ll1$,
we have defined $\tl{a}_1 \equiv a_1-\pi f_1$, and $\tilde{A}_L$ and $\tilde{A}_H$
as Eqs.~(\ref{A1}) and (\ref{A2}) with $a_1$ replaced with $\tilde{a}_1$. Here and in what follows
we drop the CP phase $\h$ in Eq.(\ref{DIV}) as well as the higher order terms represented by the dots.
We also assume the following hierarchies in the mass scales,
\begin{align}
\laq{inicon2}
H_{\rm inf} \ll M_H \ll \frac{\Lambda_1^2}{f_1}.
\end{align}
The first inequality is necessary for integrating out $A_H$ during inflation,
and the second is for the resonant conversion to take place. 

When the inflation ends, it is $\tilde{a}_1$ that is largely displaced from the 
minimum and therefore it acquires most of the initial oscillation energy. On the other hand, 
$\tilde{a}_2$ also starts to oscillate with a nonzero amplitude,
but its oscillation energy is subdominant. Therefore, for our purpose, it suffices to
study the axion dynamics in the approximated potential \eq{DMV} with 
the following initial condition,
\beq
\laq{inicon3}
\tl{a}_{1} \simeq f ,~ a_2=0,~ \dot{\tl{a}}_1=\dot{a}_2=0.
\eeq
One can of course solve the equation of motion of the two axions with the original potential,
but the above simplification enables us to capture the essence of the resonant conversion.

Let us first consider the axion dynamics soon after inflation ends. 
The $\tilde{a}_1$ oscillates in a potential \eq{DMV} which is dominated by the first term
because of the mass hierarchies \eq{inicon2} and the initial condition \eq{inicon3}, whereas
the oscillation of $\tilde{a}_2$ remains negligible for a while. Then, the axion dynamics
can be approximated by the oscillation of $\tilde{a}_1$ in a quartic potential. 
The oscillation frequency of $a_1$ is determined by the effective mass 
$m_{\rm eff}(t)$ (see Eq.~(\ref{meff})), which decreases with time. 
When it becomes comparable to $M_H$, the mixing between
the two axions become relevant, and the $A_H$ is considered to be resonantly produced.

   \begin{figure}[!t]
\begin{center}  
   \includegraphics[width=70mm]{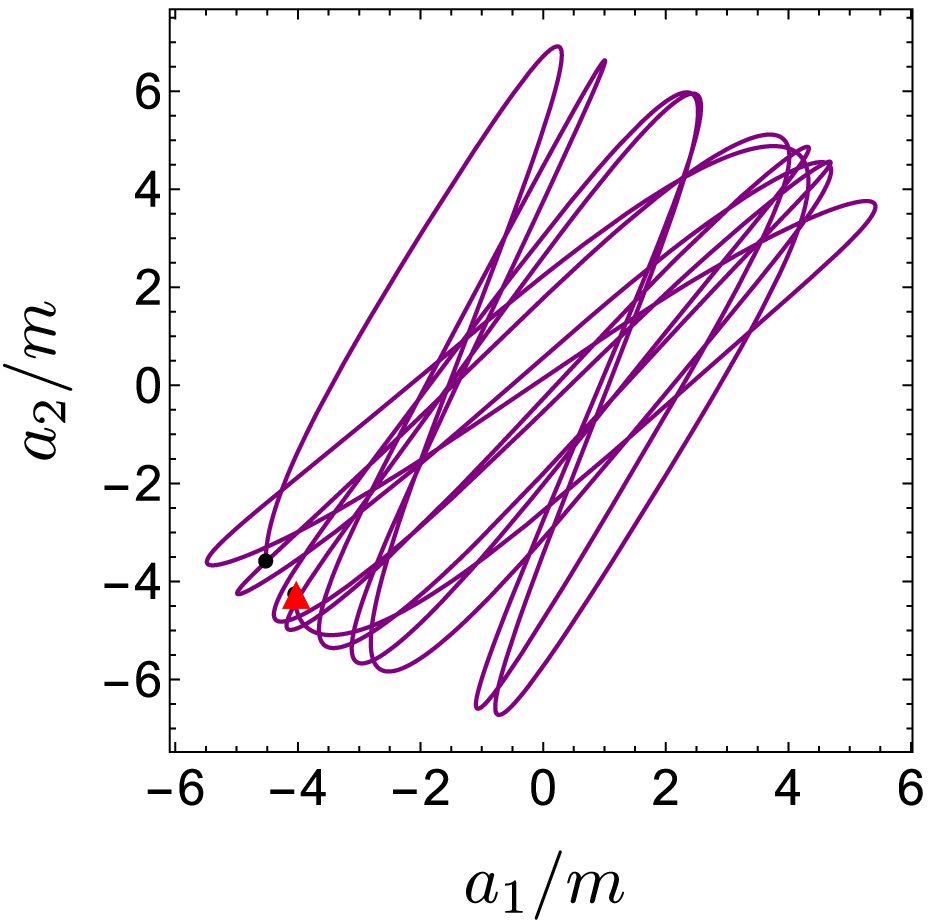}
   \includegraphics[width=72mm]{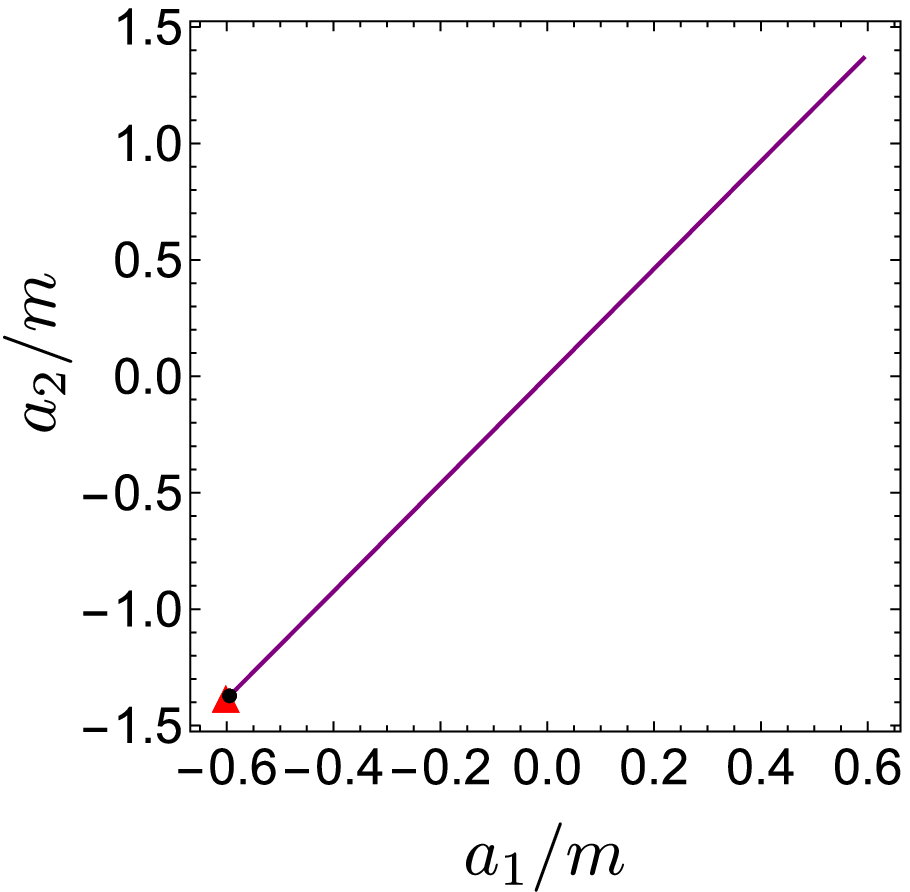}
\end{center}
\caption{ 
Typical trajectories of $a_1$ and $a_2$ when $m_{\rm eff} \sim M_H$, where 
the black dot (red triangle) represents the beginning (end) of the trajectory.
We set $\lambda' = 10^{-13}$, and  the left panel is for  $|b|=0.35$ and the right one
is for $|b|=0.43$. 
}
\laf{RC}
\laf{lc}
\end{figure}

We find two cases of the resonant conversion depending on the parameter 
 $b$ in \Eq{DMV}.  One is a chaotic conversion where  the system exhibits a chaotic behavior and
both $A_L$ and $A_H$ are excited with a similar amount. A typical trajectory when $m_{\rm eff}$ 
is comparable to $M_H$ is shown in the left panel of \Fig{RC}.  The other is an adiabatic conversion,
where the direction of the oscillations gradually change as shown in the right panel.  
This resembles the MSW effect in neutrino physics. Such a clear transition of the behavior is typical
to the chaotic system. 
In a realistic situation, however, the tachyonic preheating quickly destroys the homogeneous condensate and
it is unlikely that the adiabatic conversion takes place as the dynamics is strongly perturbed by the
spatial inhomogeneities. Instead,  the chaotic conversion could take place when the oscillation frequencies become comparable.  Therefore it is expected that both a roughly equal amount of $A_L$ and $A_H$ are excited 
by the chaotic conversion. 

After the chaotic conversion, the energy density of $A_L$ decreases as $R(t)^{-4}$, and its dissipation is 
likely inefficient since the effective mass becomes already much smaller than the initial value. 
Therefore it is $A_H$ that needs to reheat  the Universe. Since the energy density of $A_H$ decreases
more slowly as $R(t)^{-3}$, it can dominate the Universe soon after the conversion. Then, its decay
into photons reheats the Universe, diluting the abundance of $A_L$. Alternatively, the $A_H$ may quickly 
dissipate into plasma if thermal effects become important. We leave a detailed analysis of such scenario with
chaotic conversion for future work.

\section{Discussion and Conclusions}

So far we have simply assumed that the inflation driven by a light axion takes place 
at a very early stage of the evolution of the Universe, and focused on a low-scale inflation 
with the Hubble parameter at sub-eV scales when we couple the axion to photons,
i.e., the inflaton is identified with an ALP.
Here we briefly discuss issues associated with the initial condition required for such a low-scale inflation. First of all,
in order for the inflation to take place at sub-eV scales, the Universe should not collapse by 
that time, which requires some tuning of the initial spatial curvature~(see e.g. \cite{Linde:2005ht}).
For instance, if the initial spatial 
curvature is very close to zero but it is positive, the Universe may collapse before the inflation
starts. Such initial condition problem can be evaded by assuming large-field inflation such as
chaotic~\cite{Linde:1983gd} or topological~\cite{Linde:1994hy,Linde:1994wt,Vilenkin:1994pv} 
inflation models, which sets the initial spatial curvature vanishingly small. Alternatively, one may assume 
that the Universe was trapped in one of the local minima with a positive energy, experienced an old inflation, and then
a bubble was created through quantum tunneling to an adjacent vacuum with a lower energy. If our
Universe is inside such a bubble, it is almost spatially homogeneous because of the O(4) symmetric
configuration, and its initial spatial curvature is negative which prevents the Universe from collapsing.

Secondly, a priori, there is no special reason for the inflaton to sit in the vicinity of the local maximum,
since there is no enhanced symmetry at the local maximum due to the shift symmetry. While there may well be an anthropic explanation for the initial position of the inflaton, there may be another way out in relation with the bubble nucleation. 
The quantum tunneling takes place along one (or a few) of the axions, $a$,  in the axion landscape. 
Depending on how the tunneling proceeds, such axion may change its value by $\pi$ times its decay constant.
Then it would flip the sign of the inflaton potential if the axion is included in the cosine functions of $V_{\rm inf}$:
\beq
V(\phi, a) \to -V(\phi, a+\pi f_a) \supset V_{\rm inf}(\phi)
\eeq
Then, it may be possible that, before the tunneling,  the inflaton was heavy and stabilized 
at its potential minimum during the old inflation, but the local minimum at that time may turn 
into the local maximum (or saddle point) after the tunneling. 

Another cosmological issue is the origin of baryon asymmetry. In our scenario, the instant reheating
takes place just after inflation, and almost all the energy of the inflaton goes to thermal plasma.
The reheating temperature is expected to be of ${\cal O}(1-10)$\,TeV depending on the ALP mass
(see Eq.~(\ref{TR})). Since the reheating temperature exceeds $T_{\rm EW}$, the electroweak
baryogenesis could take place. It is interesting and suggestive that the reheating temperature derived by the
successful inflation and dark matter abundance happens to be of order the electroweak scale. 

We have considered the ALP couplings to photons and weak gauge bosons, but it is worth studying
the other interactions to leptons, quarks and gluons. If one couples naively the ALP to gluons, it would
be in a strong tension with astrophysical bounds, but it is possible to evade the constraints by considering
an alignment/clockwork structure with multiple axions, which generates a hierarchy in the couplings. 
In this case, it may be possible that the ALP dissipate its energy through a coupling to gluons after inflation,
but the coupling to gluons is suppressed in the present vacuum, satisfying the astrophysical bounds. 
Also it is interesting to consider a mixing with the QCD axion and the ALP(s)~\cite{Daido:2015bva,Daido:2015cba}.

The inflaton potential (\ref{DIV}) may receive additional contributions, if the shift symmetry is broken
by Planck-suppressed operators.  To be concrete, let us assume that the axion $\phi$ resides in the phase of a single complex scalar field, $\Phi$, as $\Phi = (f/\sqrt{2}) e^{i\phi/f}$. Then, one could write down 
higher-dimensional Planck suppressed operators like
\begin{align}
\frac{\Phi^\ell \Phi^{\dag (n-\ell)}}{\ell! (n-\ell)! M_{pl}^{n-4}},
\end{align}
where $n (>4)$ and $0\leq \ell \leq n$ are integers. Such Planck-suppressed operators, if exist, could affect the
inflaton dynamics.  For $f \sim 10^7$\,GeV, all the higher dimensional operators with $n \geq 7$ have negligible
impact on the inflaton dynamics or dark matter physics. One could suppress the dimension 5 and 6 operators
by imposing a $Z_N$  on $\Phi$ where $N \geq 7$. Alternatively, one of the dimension $5$ operator, e.g. $\Phi^5/M_{pl} + {\rm h.c.}$, could play a role of the second cosine function in (\ref{DIV}), because the potential height is close to
the typical value of $\Lambda = {\cal O}(1-10)$\,TeV. Then, one may impose $Z_5$ on $\Phi$ to ensure the required
potential. In principle, it may be possible that the three kinds of dimension $5$ operators, $|\Phi|^4\Phi$, 
$|\Phi|^2 \Phi^3$, $\Phi^5$, and their conjugates can generate the upside-down symmetric inflaton potential, because
they produce a cosine term with $n$ being an odd integer. The only issue of this possibility is how to suppress the
dangerous dimension $6$ operators, which may require  specific UV completion. We also note that, 
as pointed out in Ref.~\cite{Higaki:2016yqk}, the quality of the shift symmetry is improved if one considers an alignment/clockwork mechanism, because the actual breaking scale can be much smaller than the decay constant,
suppressing dangerous higher dimensional operators. 

In this paper we have proposed a scenario where inflation and dark matter are described by a single
axion in a unified manner. In particular, the inflation is driven by a light axion with a flat-top {\it and} flat-bottomed potential, which makes the inflaton practically stable on a cosmological time scale. In other words, the flatness required for successful inflation ensures the (quasi)stability of dark matter. Such a special feature of the potential is partly due to the unbroken discrete shift symmetry of the axion potential. For successful reheating, we have considered two possibilities. One is that the inflaton is coupled to photons, and the reheating proceeds mainly through a combination of the perturbative decay into two photons and dissipation process. This minimal scenario predicts the inflaton mass of $m_\phi = {\cal O}(0.01)$~eV and the photon coupling $g_{\phi \gamma \gamma} = {\cal O}(10^{-11})$\,GeV$^{-1}$, which are 
within the reach of IAXO {and the high-intensity laser experiments in the future.} 
 Also, our scenario naturally predicts hot dark matter composed of thermalized ALP whose
abundance is given by $\Delta N_{\rm eff} \simeq 0.03$. Such hot dark matter component can
be probed by the future CMB experiments such as PIXIE and CMB-S4, 
because the ALP is almost thermalized after inflation.  The other is that the inflaton gets mixed with another heavy axion, which is 
produced after inflation like the waterfall field in hybrid inflation or resonant processes. Interestingly, one of the axions can be coupled to photons 
more strongly than the light ALP in the low energy, which enables more efficient dissipation.
If the reheating proceeds through decay of the heavy axion into two photons, the ALP mass can be heavier, and indirect dark matter 
search may be useful to probe the decaying ALP dark matter. 
Furthermore, the heavy axion(s) may also be searched for by the beam dump experiments like SHiP. Further analysis along this line such as couplings of the axions to gluons or fermions, as well as the chaotic conversion based on the full potential and tachyonic preheating effects, are certainly warranted.

\section*{Acknowledgments}
We thank K. Nakayama and K. Mukaida for explaining the dissipation processes, and
K. Hamaguchi and M. Takada for useful comments. FT also thanks the organizers and participants of
CERN-EPFL-Korea Theory Institute``New Physics at the Intensity Frontier" for useful feedbacks. 
This work is supported by JSPS KAKENHI Grant Numbers JP15H05889, JP15K21733, JP16H06490, 
JP26247042,  JP26287039 (F.T.), by Tohoku University Division for Interdisciplinary Advanced Research and Education (R.D.),  and by World Premier International Research 
Center Initiative (WPI Initiative), MEXT, Japan (F.T.).

  \end{document}